\documentclass[vecphys,natbib]{svmult}

\usepackage{graphicx}        
\usepackage{multicol}        
\usepackage{times}
\usepackage[bottom]{footmisc}

\newcommand{\zz}{\mathbb{Z}}

\newcommand{\cc}{\mathbb{C}}
\newcommand{\me}{\mathrm{e}}

\usepackage{amssymb,amsmath,amsfonts,bbm}
\usepackage{wasysym}

\usepackage{psfrag}
\usepackage{ifthen}

\newcommand{\mybw}{0}		

\ifthenelse{\equal{\mybw}{1}}{\graphicspath{{figures_bw/},{./}}}{\graphicspath{{},{./}}}
\ifthenelse{\equal{\mybw}{1}}%
  {\newcommand{\bwtext}[2]{#1}}%
  {\newcommand{\bwtext}[2]{#2}}

\newcommand{\twofigures}[6]{ \begin{figure}[tb]
  \vbox{
    \hbox to \hsize{
      \begin{minipage}[b]{.47\textwidth}
        \center{\includegraphics[width=\textwidth]{#1}}
      \end{minipage}
      \hfill
      \begin{minipage}[b]{.47\textwidth}
        \center{\includegraphics[width=\textwidth]{#4}}
      \end{minipage}
    }
    \hbox to \hsize{
      \parbox[t]{.47\textwidth}{\caption{#2 \label{fig:#3} }}
      \hfill
      \parbox[t]{.47\textwidth}{\caption{#5 \label{fig:#6} }}
    }
  }
\end{figure} }

\begin{document}

\title*{Percolation in quantum computation and communication}
\author{K.~Kieling\inst{1}\inst{2}\inst{3}\and
J.~Eisert\inst{1}\inst{2}}
\institute{
QOLS, Blackett Laboratory, Imperial College London, Prince Consort Road, London SW7 2BW, UK
\and
Institute for Mathematical Sciences, Imperial College London, Prince's Gate, London SW7 2PG, UK
\and
\texttt{konrad.kieling@imperial.ac.uk}
}
\maketitle

\section{Introduction}
\subsection{Percolation theory}
Percolation theory is concerned with the behavior of connected 
clusters in a random graph. Originally developed to 
mathematically describe random media, it has over the years
matured to a field of research in its own right.
At its core is the notion of a percolation threshold: Say, on a cubic
two-dimensional lattice $\zz^2$ neighboring
vertices are connected by an edge with probability
$p$, the distributions being identical for all edges and independent. 
We say then that this edge is ``open''. Clearly,
when $p=0$, all vertices are unconnected to each other. 
If we now increase the probability $p$ starting from zero, 
more and more vertices will be connected. For small
values of $p$, this will typically lead to isolated, 
non-crossing, small clusters of connected vertices. 
With increasing probability, the size of the clusters will 
also increase. The key observation now is that there
exists a {\it critical percolation threshold} $p^{\rm (c)}_{2}$ 
that marks the arrival of an infinite crossing cluster.
For $p<p^{\rm (c)}_{2}$, all open clusters are finite, whereas
for $p>p^{\rm (c)}_{2}$ there exists almost certainly an open infinite
cluster. This behavior of having a
connected open crossing cluster in a finite subset of 
$\zz^2$ can also be impressively documented by simple
numerical simulations. Since the early work on 
fluid flow in a porous stone giving rise to the first simple
percolation model \cite{Hammersley}, percolation 
theory has found numerous applications in 
a quite impressive portfolio of diverse fields, besides
being a mathematical discipline in its own right
\cite{Grimmett}: 
This includes 
the study of disordered many body systems in classical and 
quantum physics, of instances of fluid dynamics and fire 
propagation, 
of reliability of processes, or of many aspects 
of statistical mechanics. 

When it comes to applying percolation theory to physical
systems, this article forms an exception in this book. This 
chapter is concerned with the {\it connection 
between percolation theory on the one hand and 
quantum information processing on the other hand}. 
We are facing a percolation-type situation somewhat
similar to the one encountered in the study of disordered
systems.
\begin{itemize}

\item However, in the context of quantum information processing,
the {\it randomness} we encounter is not the one of 
classical probabilistic parameters in a model, but 
rather originates
from the statistical
character of quantum theory under measurements as such.

\item Also, it is not the aim to model physical systems in
their unknown properties, but rather think of {\it engineered 
quantum systems} that have been prepared in a very specific
state for a very specific purpose. In this sense, the 
emphasis is less on {\it a posteriori}
description of properties of systems
or materials that are already there. But instead, to explore
what state preparation and what communication protocols
are possible, when limited to probabilistic processes
in quantum information applications.
\end{itemize}
This chapter forms an instance of a ``review'' on the 
link between percolation theory and quantum information
theory, as a topic that is presently receiving notable 
attention. We will see that this link is a quite natural one, 
where percolation in several ramifications
enters as a concept to overcome limitations imposed by 
probabilistic processes that 
occur in measurement processes. Yet, whereas
this link has been mentioned a number of times in
the literature, so far, only four research articles flesh 
out this link in detail \cite{KRE06,ACL07,PCAL07,BEF+07}. 
This article can hence 
be seen as (i) a short review on the material  
that is known and published at present, as (ii) a presentation of
a long and more detailed version of the short proof
presented by the same authors in Ref.\ \cite{KRE06}, 
and, most importantly, as an (iii)  invitation to the subject.
Readers familiar with elements of quantum information
theory may wish to skip to 
Section \ref{PercolationComputing}.

\subsection{Quantum computation and communication}

{\it Quantum information processing} takes the idea seriously
that  when storing or processing information, it matters
whether the underlying physical system follows classical
or quantum laws. In classical information theory, one is
used to the fact that it hardly makes sense to think
of the physical carrier of information, as one can 
transform the information stored in one form to another
carrier in a lossless fashion anyway. This abstraction from the
physical carrier is challenged when one thinks of {\it single
quantum systems} forming the elementary processing
units. Indeed, the task of transforming
the ``information stored in a quantum system'' into 
classical information and back is 
impossible. Quantum information processing is however
not so much concerned with limitations due to quantum
effects, but rather thinks of new applications in 
computing and communication when the carriers
of information are single quantum systems.

A \emph{quantum computer} \cite{nielsen,EW04} is 
such an envisioned device: 
One thinks of having an array of quantum systems -- spins, say,
referred to as {\it qubits}. 
This system,
associated with a Hilbert space ${\cal H}=(\cc^2)^{\otimes n}$,
is initially prepared
in a known, preferably pure quantum state described by a 
state vector $|\psi\rangle$. Then one 
manipulates the state by means of unitary time evolution or
by means of measurements. Acknowledging that not
any unitary dynamics is accessible on a quantum
many-body system, in the {\it circuit model}, this step
of computation is broken down to {\it quantum gates}  \cite{nielsen,EW04}.
One hence implements a sequence of unitary gates
that have trivial support on all sites except single sites -- giving
rise to {\it single qubit gates} -- and pairs of sites -- {\it two-qubit
gates}. The state vector 
after the unitary time evolution is then
\begin{equation}
	|\psi\rangle = \prod_j U_j |\psi\rangle.
\end{equation}
This is followed by local individual measurements on the spins.
The measurement outcomes at the end then deliver 
(typically statistical) data, from which
the outcome of the computation can be estimated. The
important point is that because the quantum system follow
the laws of quantum mechanics and can be prepared in a
superposition exploring an exponentially large Hilbert space,
some problems could be solved with significantly less effort
on this envisioned device than on any classical computer. In fact, 
they could be solved with polynomial effort in case of 
some problems like factoring 
that are believed to be computationally 
hard classically.

Yet, the circuit model is only one out of several models 
for quantum computation. In the computational model
having center stage in this chapter, one abandons
the need for unitary control via quantum gates, and 
performs only local measurements on an entangled
quantum state of many constituents. This is the 
model of the {\it one-way computer} \cite{RB01,BR01,HDE+06}. 
The  entanglement  present in the initial resource then
facilitates the efficient simulation of any other 
computation that had unitary gates available. Needless
to say, in any such approach, the core question is how
to actually prepare this initial resource state, an instance of a
{\it cluster} or {\it graph state} \cite{HEB04,HDE+06}. 
It is in this 
preparation -- and in the way to overcome the intrinsically
probabilistic character of quantum measurements -- that
percolation ideas will be important in the first part of
this chapter. It will turn out that by making use of 
percolation ideas, some of the key obstacles in some
physical architectures to realize quantum computation can
be weakened. In particular, in optical approaches, we will
see that percolation ideas facilitate the preparation of
such states with little dependence of preparation steps on
earlier measurement outcomes -- an experimentally very
difficult prescription in this context. Percolation will hence
help us to think of what is known as 
linear optical quantum computation 
\cite{KLM01,WRR+05,KRE06,BEF+07}.

The second part of this chapter is dedicated to a related
but different question. This is taken from {\it quantum 
communication} and {\it key distribution}: 
Several applications in quantum information processing,
most importantly, quantum key distribution for the secure
transmission of classical data, but also instances of 
the envisioned distributed quantum computing, rely on the
availability of almost maximally entangled states shared between
spatially separated laboratories. If one has several such stations
on the way, each communicating and exchanging quantum
particles with a number of neighbors, one arrives at the idea
of a {\it quantum network}. In Section \ref{section:networks} we will
review the use of percolation ideas to establish 
maximally entangled pairs for quantum communication
between arbitrary stations, based on the published work
Ref.\ \cite{ACL07,PCAL07}. 
Here, the main point will be that not only
classical percolation ideas can be employed to meet the
aim of sharing entanglement between arbitrary vertices of a 
graph: Instead, appropriate quantum measurements allow
to outperform the situation of classical edge percolation. 
In this sense, one encounters a notion of
{\it entanglement percolation}. In fact, such entanglement percolation
can drastically outperform strategies based on 
standard edge percolation.

\section{Percolation and quantum computing}\label{PercolationComputing}
\subsection{General concepts}

We now turn to the preparation of cluster or more generally
graph states as resource state for measurement-based quantum computation.
This measurement-based model is a desirable route to quantum computation
in a number of architectures. This is not the least due to the very clear
distinction between the preparation of entanglement, and the consumption 
thereof by means of measurements, abandoning the need for unitary
gates. Percolation ideas play the crucial role 
here when thinking 
of errors, of dealing with probabilistic gates building up the states
and, most importantly, of coping with the daunting feedforwards
(dependence of later action on earlier measurement outcomes).

It will be necessary -- before we establish the percolation argument -- 
to introduce a number of concepts that will be used later on.
A {\it graph state} \cite{SW02,HEB04,NDM04}
is a quantum state
defined on lattice that is described by 
an undirected graph $G=(V,E)$, with $V$ being the set of vertices,
and $E$ the set of edges. The vertices are embodied by
physical systems, so single spins or qubits. The edges
represent interactions. More specifically, a
{\it graph state} is the  
simultaneous eigenstate to the set of \emph{stabilizers}
\begin{equation}
  K^{(a)}_G = \sigma_x^{(a)} \bigotimes\limits_{b:{\text{dist}}(a,b)=1} \sigma_z^{(b)}
  \label{eqn:stab}
\end{equation}
for all $a\in V$ with eigenvalue $1$, i.e.,
\begin{equation}
  K^{(a)}_G|\psi_G\rangle = |\psi_G\rangle \qquad \forall a\in V .
\end{equation}
Here, ${\text{dist}}(x,y)$ is the graph-theoretical distance 
\cite{BondyMurty} between vertices $x$ and $y$ on $G$, i.e., 
${\text{dist}}(x,y)=1$ for neighboring vertices. Here and in the following,
$\sigma_{x,y,z}^{(b)}$ denotes a
Pauli operator with support on the 
Hilbert space of the physical system labeled $b$.

Equivalently, this state
can be thought of as having each qubit prepared in the state vector
$|+\rangle= (|0\rangle + |1\rangle )/\sqrt{2}$ 
and applying an interaction leading to a {\it phase gate}
or {\it controlled-Z}-gate 
\begin{equation}
	U_z= |0,0\rangle\langle0,0|
	+ |0,1\rangle\langle0,1| + |1,0\rangle\langle1,0| - 
	|1,1\rangle\langle1,1|
  \label{eqn:gate}
\end{equation}
to neighboring vertices $a,b\in G$, so to vertices that are connected by an edge $(a,b)\in E$.
Because these gates are diagonal in the computational
basis, they commute, and hence
the order in which they are applied does 
not influence the final state. 

Such a graph state may be defined for any graph. Most important for our purposes are lattices where $V\subset \zz^2$,
so finite qubic lattices in 
two dimensions. 
Such a graph state of a cubic lattice is called a 
\emph{cluster state} \cite{BR01}. It has been
shown that local measurements on single constituents of such a 
cluster state are just as powerful as the gate model for quantum 
computation, and it can hence efficiently simulate any other
quantum computer \cite{RB01,RBB03}. That is, 
for a given finite computation there exists a finite graph
with $V\subset\zz^2$ such that the computation
can be simulated by a sequence 
of suitable single-qubit local  measurements on the cluster state.
The usual mapping from a circuit model to a cluster state allows to label one dimension
as time which comes in handy because only time slices of a cluster actually
have to exist at a given point during the computation.
We will also frequently encounter 
finite subsets of the hexagonal lattice which is a computation resource in the same sense.
Note that {\it a priori} the lattice does not correspond to any spatial lattice --
what determines the quantum state are solely the adjacencies on the graph.

When preparing cluster states, there are two main types of 
errors where percolation comes into play:
\begin{itemize}
\item One can think of the above lattice structure emerging from a 
{\it physical lattice}. This may be an {\it optical lattice} generated
by standing wave laser light, where atoms are
located at individual lattice sites \cite{MGW+03}.\footnote{We 
leave issues that are under significant consideration in the 
literature such as the question
of locally addressing single sites in measurements aside and 
focus on the
preparation of the cluster state.} 
By means of interactions or controlled collisions, implementing (\ref{eqn:stab}) or (\ref{eqn:gate}), respectively, a 
cluster state can in principle be
prepared in such systems. 
Yet, Mott hole defects, where
sites are left unoccupied in a random fashion, lead to defects.
This is a situation where ideas of
{\it site percolation} can overcome the problem of having 
non-unit filling factors.

\item Using other types of quantum systems, cluster states have to be prepared sequentially, by means
of explicitly employing gates to pairs of constituents. This should not be confused
with the absence of gates when doing the actual computation via
measurements -- the gates used during state preparation are independent of the actual algorithm to perform.
The most resource-efficient and hence most
feasible ideas of doing quantum computation with 
{\it linear optical systems} or with {\it atoms in optical
cavities} (see, e.g., Ref.\ \cite{LBB+06} and references therein)
rely on exactly such a sequential build-up of 
the cluster state. Here, one starts from elementary 
building blocks, like entangled photon pairs. From these
building blocks the full cluster or graph state can be 
prepared by means of a sequential 
application of controlled-$Z$ gates. During the course of this process,
in principle gates between any pair of qubits are allowed, which one is actually
applied at a given step depends on a chosen prescription which, in turn,
depends on outcomes of earlier gate applications.
The lattice structure is imposed by the sequence of gates, so every state described by
an undirected graph without degenerate edges can be generated.

This being the most promising approach to think of optical 
quantum computing, it seems important to address two
major challenges that have to be overcome in such an
approach:\footnote{Again, we will focus on the intrinsic
problems of such an idea. From a physical perspective,
the development of heralded sources with high efficiency
and mode quality, as well as detectors with high detector
efficiency are major obstacles, a significant research
effort is dedicated to.}.

\begin{enumerate}
\item The applied gates necessarily 
operate in a probabilistic fashion.\footnote{This is a consequence of
gates being in turn implemented by means of measurements.
All non-linearities in {\it fully linear optical systems} have
to be effectively realized by means of measurements, and this
randomness is hence the intrinsic randomness in measurements
in quantum mechanics. Gates like the 
\emph{fusion gates}~\cite{PJF01,BR04} act in effect
as the desired controlled-$Z$ gate, albeit in a probabilistic
fashion. In approaches
based on {\it atoms in optical cavities, coupled via light}
similar issues occur.} Indeed, the probability of success
of such gates is typically quite small \cite{Eisert04,SL04}. Hence,
when sequentially preparing a cluster state, needless to say,
the very order in which the states are sent through the entangling gates plays a crucial role~\cite{KGE06,GKE06,KGE07}. Especially the success 
probability of the fusion gates, $p_{\rm success}=1/2$, will be of central interest as this is already the optimal probability 
for such a gate in linear optics \cite{CL01,GKE06}. 
A naive approach of trying to grow an
$n\times n$ cluster state by using such gates will fail. 
However, this obstacle can in principle be overcome: One can 
show, using combinatorical methods and ideas of
convex optimization, that there exist methods to achieve
an optimal scaling of $O(n^2)$ invocations of quantum gates (and number of qubits),
despite the gates operating probabilistically \cite{GKE06}. 

\item Unfortunately, this -- and actually any such -- 
procedure leads to a great deal of conditional dynamics: Depending on
outcomes of earlier fusion operations, one has to decide which 
pair of states will be used in the next step. This, however, turns 
out to be a significant challenge in actual experimental realizations,
as it requires
active switching including
coherent interaction between any pair of modes in the setup,
while unused modes are stored in quantum
memories, thus clearly rendering it a very difficult prescription.
\end{enumerate}

To lessen this daunting requirement, percolation ideas can
come into play: One could think of a static setup, 
i.e., an underlying structure is imposed onto the resources
and only applications of entangling gates
between nearest neighbors are allowed. The probabilistic
nature of quantum gates is the source of randomness,
we now face a problem of {\it edge percolation}. How exactly this
will work and how to find bounds to the scaling of the 
resource requirements to prepare two-dimensional cluster
states even if the probability of an edge being open 
is smaller than the critical percolation threshold 
for a two-dimensional cubic lattice $p<p^{\rm(c)}_{2}$,  
will be the subject of the next section
\cite{KRE06}. The latter property -- which is responsible
for some technicalities -- is important, as the probability
of success of the fusion gate happens to be 
$p_{\rm success}=1/2$, which is identical with the 
critical percolation threshold $p^{\rm(c)}_{2}=1/2$. Hence,
ways have to be found to achieve percolation even using
such gates.
\end{itemize}

Further developing these ideas, one can ask whether there
is a  \emph{phase transition} in edge percolation
with respect to quantum computing applications: One thinks
of a cubic lattice as in 2., with edges being open in case a
gate was successful. If $p>p^{\rm(c)}_{2}$, one can
extract a resource almost certainly that allows for universal
quantum computing. Below the threshold $p<p^{\rm(c)}_{2}$, 
one can show that one can almost certainly simulate the 
evolution of the quantum system on a classical computer. 
Hence, this regime is not only not useful for quantum 
computation, but even classically efficiently 
tractable.\footnote{A similar
situation has been observed in Ref.\ \cite{GESP07}, where a
{\it Kitaev's toric code state} is not a universal resource, but can
be classically efficiently simulated \cite{BR07}. 
If one modifies the state
very little by means of local phases, one can not only not
keep track of measurements classically in an efficient fashion.
But in fact, it can be shown that the state would serve as a
universal resource for quantum 
computation \cite{GE07,GESP07}.} In this sense, there is
a phase transition in the computational potency of the
resource depending on the probability of an edge being open.

In these different flavors we will encounter percolation ideas
in the context of quantum computing. To be brief, we 
concentrate on the latter two aspects, putting an emphasis
on bond percolation (so all $p$ will be bond probabilities
unless stated otherwise). We note, however, that the problem
of site percolation can be treated similarly.

\subsection{Resource state preparation in measurement-based computing} \label{section:stateprep}

\subsubsection{Renormalization}
The goal of this section will be to show how to generate a cluster state of a given size with probabilistic entangling gates (succeeding with probability $p$)
almost certainly with the help of bond percolation.
The lattice will be divided into blocks which will be reduced later to single qubits, thus ``renormalizing'' the lattice.
Whenever a block contains a crossing open cluster that connects the block's four faces (in the first two dimensions) we will refer of this crossing cluster as a renormalized qubit.
If the crossing clusters of two neighboring renormalized qubits acually touch each other (i.e., there exists an open path between two vertices of the first and the second renormalized qubits
that lies completely within the union of the two respective blocks, and they are connected by open bonds),
then the reduction of the blocks to single qubits will yield a ``renormalized bond'' between these qubits.
How this reduction actually works is the scope 
of the following sections.
Now, we are looking for the probability $P_p(\mathfrak U(L,k))$ of the event $\mathfrak U(L,k)$ to occur.
$\mathfrak U(L,k)$ denotes the event that the renormalized square lattice of size $L\times L$ with hypercubic blocks of size $k^{\times d}$
is fully occupied and connected. Given a dependence of the block on the lattice size, $k(L)$, we will use the abbreviation
$P_p(L)=P_p(\mathfrak{U}(L,k(L)))$.

The result is more precisely phrased as follows:
\begin{theorem}[Resource consumption] \label{thm:scaling}
  Let $p>p^{\rm(c)}_{d}$, $d\ge2$ being the dimension of a
  hypercubic lattice. 
  Then for any $\mu>0$, the probability 
  $P_p(L)$ of having an $L\times L$ renormalized
  cubic lattice fulfills
  \begin{equation}
 \lim\limits_{L\rightarrow\infty}P_p(L) = 1, \end{equation}
  with an overall resource consumption of $R(L)=O(L^{2+\mu})$.
\end{theorem}
Here, $p^{(\rm c)}_{d}$ denotes the critical percolation threshold 
of the $d$-dimensional hypercubic lattice.
$R$ is used to refer to the number of initial resources, so the constituents that are placed on each lattice site with the ability of
``growing'' connections to their neighbors in a probabilistic manner.
Note, that the dimension $d$ can always be chosen such that the gates at hand ($p>0$) operate in a regime that allows for percolation ($p>p^{\rm(c)}_{d}$).

\begin{proof}
\noindent\emph{($d\ge 3$)} In the case of dimensions $d>2$,
so different from $d=2$, 
crossing paths in different directions not necessarily intersect. 
This approach is often nevertheless favorable due to the
higher critical percolation threshold in higher dimensions.
Let us fix $L\in\mathbb{N}$ and take 
\begin{equation}
	U = [1,2kL]^{\times2}\times[1,2k]^{\times d-2} 
	\subset\mathbb{Z}^d
\end{equation}
for some $k\in\mathbb{N}$. This slab can be divided
into $L^2$ disjoint hypercubes with an edge length of $2k$. With $A_y(k)$, $y=(y_1,y_2)\in [2,2L]^{\times2}$ we denote the $(2k)^{\times d}$ hypercube starting at $(y_1k,y_2k,1,\ldots,1)$.
For $y=2x$, 
\begin{equation}
	x\in M=[1,L]^{\times 2}, 
\end{equation}
these hypercubes $A_y(k)$ are the disjoint blocks, and $M$ plays the role of the renormalized square lattice.

Furthermore, we will use the overlap between adjacent blocks in the first direction, 
\begin{equation}
	B_{y}(k)=A_{y}(k)\cap A_{(y_1+1,y_2)}(k) 
\end{equation}
for $y_1=2,\ldots,2L-1$, and the union of disjoint neighboring blocks
in the second direction, $C_{z}(k)=A_{z}(k)\cap A_{(z_1,z_2+2)}(k)$, $z_2 = 2,4,\ldots,2(L-1)$.

On these blocks we will define a series of {\it events} as follows:
\begin{itemize}
  \item $\mathfrak{A}_y(k)$: There exists an open left-to-right crossing cluster in $A_y(k)$ in the first dimension, so an open path containing open vertices $a$ and $b$ with
    $a_1=y_1k + 1$, $b_1=(y_1+2)k$.
    For $p>p^{\rm(c)}_{d}$ there exists a constant $g>0$, only dependent on $p$, such that \cite{Grimmett}
      \begin{equation} \label{eqn:block}
 	P_p(\mathfrak{A}_y(k)) \ge 1-\exp(-gk^2) .
      \end{equation}

  \item $\mathfrak{B}_y(k)$: The number of open left-to-right crossing clusters in $B_y(k)$ (see Fig.~\ref{fig:eventB}) does not exceed $1$.
    It is shown in Ref.~\cite{Aizenman97} that for $p>p^{\rm(c)}_{d}$ there exist constants
    $a,c>0$, only dependent on $p$, such that the probability of $\mathfrak{B}_y(k)$ occuring satisfies
    \begin{equation}
 P_p(\mathfrak{B}_y(k)) \ge 1-(2k)^{2d}a\exp(-ck) .
 \end{equation}

  \twofigures{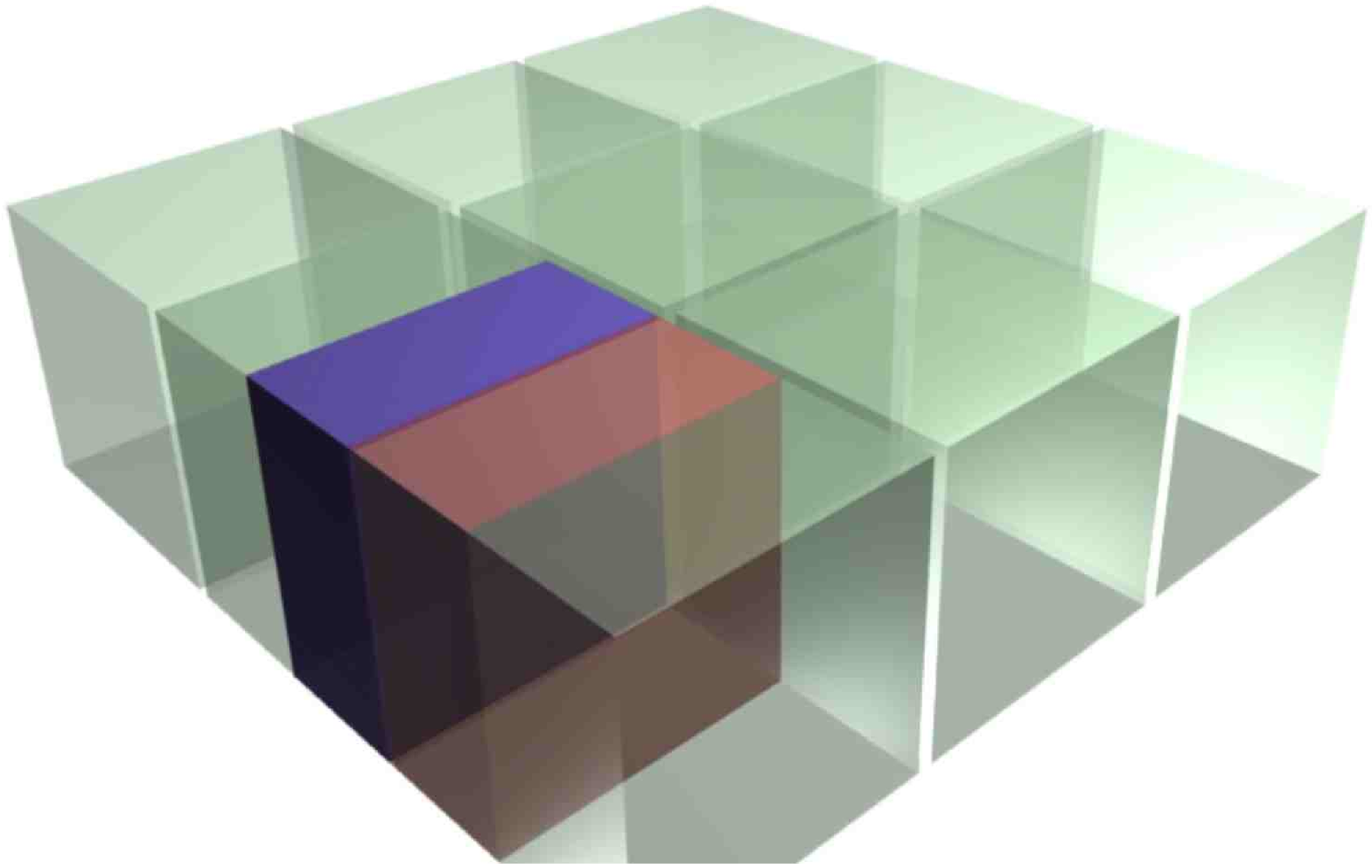}{The blocks $A_y$ for $y=2x$. The \bwtext{two highlighted regions}{red and blue regions together}
               constitute the block $A_{(3,2)}$, and the \bwtext{darkest one}{blue region} is an overlap with a neighboring block, so $B_{(3,2)}$.}{eventB}
             {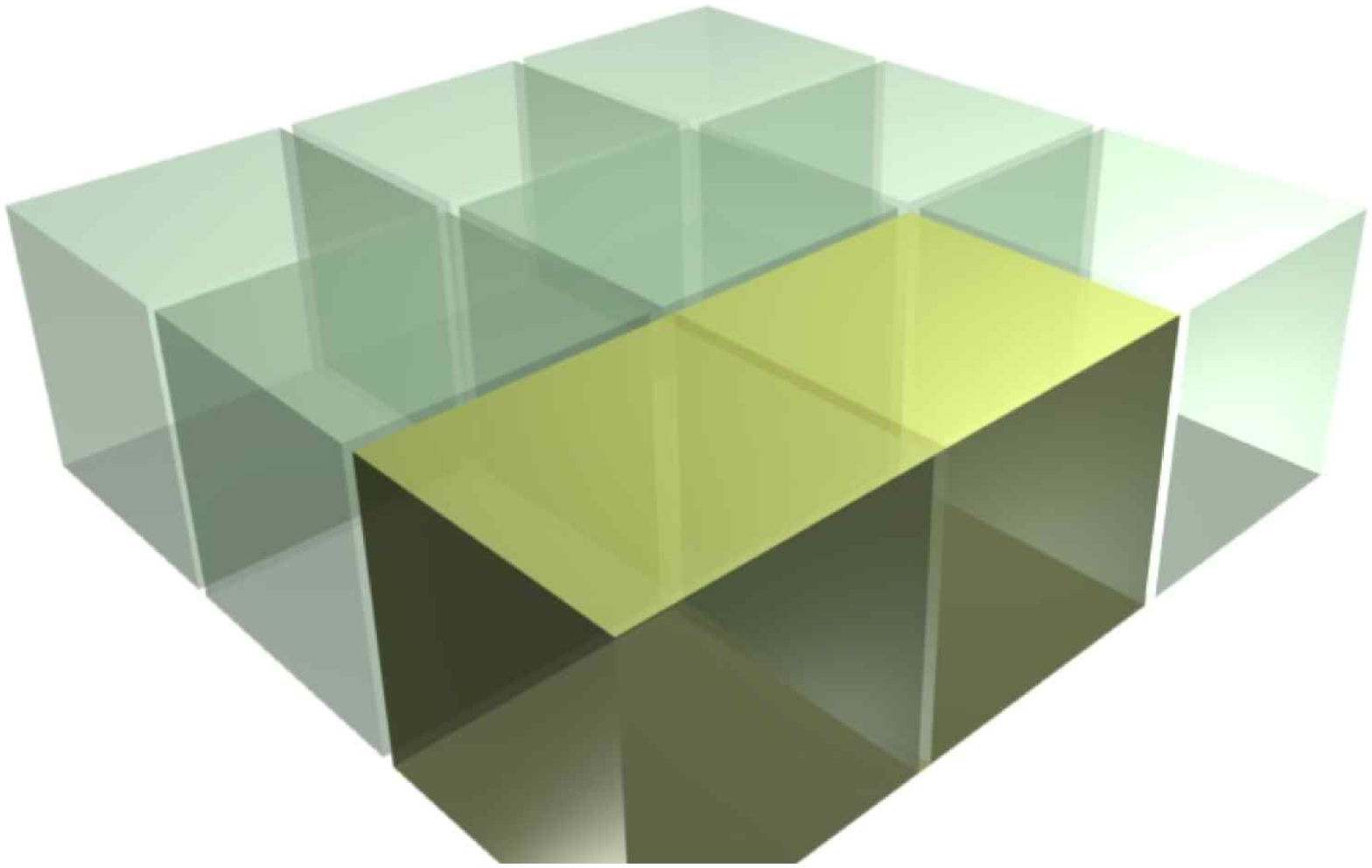}{Again, the blocks $A_y$ for $y=2x$ 
               are shown and the \bwtext{highlighted}{yellow} region is the intersection $C_{(2,2)}$.}{eventD}

  \item $\mathfrak{D}_y(k)$: We can now make use of the two events defined before. Let $\mathfrak{D}_z(k)$, $z_1=3,5,\ldots,2L-1$, $z_2=2,4,\ldots,2L$, be the event that there exists an open
    left-to-right crossing cluster in both
    blocks, $(z_1-1,z_2)$ and $(z_1+1,z_2)$, and these two clusters are actually connected (see Fig.~\ref{fig:eventD}). That means, there exists an open left-to-right crossing cluster in
    $A_{(z_1-1,z_2)}(k)\cup A_{(z_1+1,z_2)}$. The events $\mathfrak{A}_y(k)$ and $\mathfrak{B}_y(k)$ are \emph{increasing events}\footnote{Let $\chi(\mathfrak{A}_p)$ denote the characteristic
    function of the event $\mathfrak{A}$ for an elementary probability $p$. An increasing event $\mathfrak{A}$ is one that satisfies $|\chi(\mathfrak{A}_p)|\le|\chi(\mathfrak{A}_{p'})|$
    if $p\le p'$.}.
    That allows for application of the
    \begin{lemma}[FKG inequality] Let $\mathfrak{X}$ and $\mathfrak{Y}$ be increasing events. Then
    \begin{equation}
      P_p(\mathfrak{X}\cap\mathfrak{Y})\ge P_p(\mathfrak{X})P(\mathfrak{Y}) .
    \end{equation}
    \end{lemma}
    This results in an upper bound for
    $P_p(\mathfrak{D}_y(k))$, with
    \begin{equation}
    \mathfrak{D}_y(k)=\bigcap_{a=0,\pm1}\mathfrak{A}_{(y_1+a,y_2)}(k)\cap\mathfrak{B}_{(y_1-1,y_2)}(k)\cap\mathfrak{B}_{(y_1,y_2)}.
    \end{equation}

  \item $\mathfrak{E}_z(k)$: The next event we need is the one that ``connects'' two blocks in the second dimension.
    Let $\mathfrak{E}_z(k)$ be the event that there exist at most one open left-to-right crossing cluster in $C_{z}(k)$.
    In order to apply the arguments of Ref.~\cite{Aizenman97},
    we extend the blocks in the last $d-2$ dimensions by another $2k$.
    The probability of $\mathfrak{E}_z(k)$ occuring is bounded by
    \begin{equation} 
    		P_p(\mathfrak{E}_z(k)) \ge 1-(4k)^{2d}a\exp(-2ck) .
		\label{eqn:atmostone} 
    \end{equation}

  \item The last event, $\mathfrak{F}_z(k)$, we will define here is that of having an open left-to-right crossing clusters in both, $A_{(z_1,z_2)}(k)$ and $A_{(z_1,z_2+2)}(k)$, but at most
    one in $A_{(z_1,z_2)}(k)\cup A_{(z_1,z_2+2)}(k)$. That means, again, that there is actually one left-to-right crossing cluster in $C_z(k)$, but it connects the left and right
    faces of $A_{(z_1,z_2)}(k)$ and $A_{(z_1,z_2+2)}(k)$ simultaneously.
    Again, by using the FKG inequality we can construct an upper bound to the probability of occurrence of
    $\mathfrak{F}_z(k) = \mathfrak{A}_{(z_1,z_2)}(k)\cap\mathfrak{A}_{(z_1,z_2+2)}(k)\cap\mathfrak{C}_z(k)$.
\end{itemize}

Having all these events at our disposal, the goal of realizing a fully renormalized lattice can be formulated quite easily: we are looking for a simultaneous occurence of $\mathfrak{D}_z(k)$
and $\mathfrak{F}_z(k)$ for a suitable set of $z=(z_1,z_2)$'s, so
\begin{equation}
 \mathfrak{U}(L,k) = \left(\bigcap\limits_{ z_1=3,5,\ldots,2L-1 \atop z_2=2,4,\ldots,2L }\mathfrak{D}_z(k) \right)\cap
                     \left(\bigcap\limits_{ z_1=2,4,\ldots,2L \atop z_2=2,4,\ldots,2(L-1) }\mathfrak{F}_z(k) \right) . \end{equation}
By subsequent application of the FKG inequality we can express an upper bound to the probability of $\mathfrak{U}(L,k)$ occuring in terms of the probabilities
of the former events:
\begin{eqnarray}
   P_p(\mathfrak{U}(L,k)) &\ge& \prod\limits_{y_1=2,3,\ldots,2L   \atop y_2=2,4,\ldots,2L}     P_p(\mathfrak{A}_y(k))	
                                \prod\limits_{y_1=2,3,\ldots,2L-1 \atop y_2=2,4,\ldots,2L}     P_p(\mathfrak{B}_y(k))	
                        \times \nonumber \\ &&
                                \prod\limits_{y_1=2,4,\ldots,2L   \atop y_2=2,4,\ldots,2(L-1)} P_p(\mathfrak{E}_y(k))	
                                \nonumber \\
                          &\ge& (1-\exp(-gk^2))^{2L^2-L} 
                          \times \label{eqn:PU} \\
                             && ((1-(2k)^{2d}a\exp(-ck))^2
                             (1-(4k)^{2d}a\exp(-c2k)))^{L(L-1)} .  \nonumber
\end{eqnarray}

Now, we will have to find the block size scaling $k(L)$ such that this probability is approaching unity for large $L$. Moreover,
we are looking for a ``good'' scaling, in the sense that the overall resource scaling $L^2k(L)^d$ does not differ too much from the optimal $O(L^2)$.
In order to invert~(\ref{eqn:PU}), so to find the best $k(L)$ consistent with this approach, we, however, still need to relax the problem
to some extent.
By using the slowest increasing term in~(\ref{eqn:PU}) we can bound the expression from above: there exists an integer $k_0$ such that
\begin{equation}
 	P_p(\mathfrak{U}(L,k)) \ge (1-(2k)^{2d}a\exp(-ck))^{5L^2} 
 \end{equation}
for all $k\ge k_0$. Let us now use the ansatz $k=\lceil L^{\varepsilon}\rceil$, for some $\varepsilon>0$.

For any $x\in \mathbb N$ there exists a $L_0\in \mathbb N$ such that for all $L\ge L_0$
\begin{equation}
 1 - A(2L^{\varepsilon})^{2d} \exp(-cL^{\varepsilon}) \ge 1 - 1/(xL^2) . \end{equation}
Further,
\begin{equation}
 \lim_{L\rightarrow\infty} \left(1-1/(xL^2)\right)^{5L^2} = \me^{-5/x} 
 \end{equation}
and for every $\epsilon>0$ we can find an $x$ such that $1-\me^{-5/x}<1-\epsilon$.

Therefore, the chosen dependence of $k$ on $L$ is sufficient to achieve a success probability within a chosen $\epsilon$ around $1$, getting arbitrary close
in the limit of large $L$:
\begin{equation}
 \lim\limits_{L\rightarrow\infty} P_p(\mathfrak U(L,k(L))) = 1 . 
 \end{equation}
Combining this with the number of blocks used, $L^2$, this induces a resource scaling of $R(L)=O(L^{2+d\varepsilon})$.
\vspace{1em}

\noindent\emph{($d=2$)} In the two-dimensional case the connection between paths in the two directions is, of course, not an issue -- whenever a block $A_y$ is crossed in both directions, these two crossing
paths necessarily intersect. The events needed in this case are the following:
\begin{itemize}
  \item $\mathfrak{G_{i,r}}$: The rectangle $C_{i,r}=\cup_{y, y_r=i}A_y$ is crossed in the $r$-th dimension.
    Here, $C_{i,r}$ is the $i$-th row or $i$-th column in case of $r=1$ or $r=2$, respectively. The probability for such an event to happen
    satisfies~\cite{Grimmett} \begin{equation}
P_p(\mathfrak{G_{i,r}}) \ge 1 - s k L \exp( -t k ),\end{equation}
 with $s,t>0$.
\end{itemize}
Again, the $\mathfrak{C}_{i,r}$ are increasing events, so the probability of simultaneous crossing in all rows and columns,
\begin{equation}
 \mathfrak{U} = \bigcap\limits_{i=1,\ldots,L \atop r=1,2} \mathfrak{G}_{i,r} \end{equation}
satisfies
\begin{equation}
 P_p(\mathfrak{U})(L,k) \ge ( 1 - skL\exp(-tk))^{2L} .\end{equation}
Now, the choice of $k=\lceil L^{\varepsilon}\rceil$ can be used again, together with the last steps for the case $d>2$. Thus, in the two-dimensional case
a resource scaling of $R(L)=O(L^{2+d\varepsilon})$ holds as well.

\end{proof}

Although the proof was explicitly stated in terms of bond percolation, a reasoning along these lines will hold as well for site percolation or mixed site/bond percolation,
as long as the probabilities in question are above the respective threshold.

\subsubsection{Path identification} \label{section:paths}
For the cluster state to be of any use in quantum computing, the number of resources required to simulate a given quantum circuit have to depend efficiently (i.e., polynomially) on
the size of the circuit. As the size $L$ of the cluster state required to implement a given circuit has a polynomial dependence on the circuit's size \cite{RB01},
Theorem~\ref{thm:scaling} already provides a suitable scaling in the number of qubits required.

Still, the amount of time and classical memory required to implement a given computation has to obey a well-tempered scaling as well.
The ``quantum part'' (i.e., the number of subsequent measurements in the preparation- and in the computing stage) 
only requires $O(1)$ of time for preparation of the initial pieces, a single step for all the simultaneous entangling operations,
and the measurements to reduce the cluster to the renormalized one and perform the computation. Many of them can be performed in parallel,
but an upper bound is given by $R(L)$.

The classical amount of memory, of course, starts with $R(L)$ to store all gate outcomes (and therefore the percolated graph).
In the following, the scaling of classical resources 
will be 
analyzed \footnote{More details and a 
MatLab implementation of the relevant parts can be found in
the supplementary material of Ref.\ \cite{KRE06} at
http://www.imperial.ac.uk/quantuminformation. }.

\paragraph{Crossing clusters}
For identification of the crossing clusters within the blocks,
cluster finding algorithms like the Hoshen-Kopelman-algorithm \cite{HK76} can be employed.
Out of the box this would require $O(k^{d-1})$ of classical memory and $O(k L^2)$ timesteps.
If there exist more than one crossing cluster (which is as of~(\ref{eqn:atmostone}) highly improbable), only a single one 
(e.g., the one with the largest surface) will be chosen
for the subsequent procedure.

\paragraph{Connecting the blocks}
A ``mid-qubit'' which is a member of a crossing cluster near the center of the block will be chosen ($R(L)$ time-steps) in every block.
Let us define an open path on $G=(V,E)$ between $a_1,a_{n+1}\in V$ by
$\mathcal P(a_1,a_{n+1})=\{(a_1,a_2),(a_2,a_3),\ldots,(a_n,a_{n+1})\}$ with $(a_i,a_i+1)\in E$, $i=1,\ldots,n$, and
its length by $|\mathcal P(a_1,a_{n+1})|=n$. Because we are using undirected graphs, $(a,b)=(b,a)$ and there is a corresponding path from $B$ to $a$ for each path from $a$ to $b$.
Open paths between the mid-qubits of all pairs of neighboring blocks are identified using a breadth-first-search (BFS) algorithm \cite{sedgewick} ($R(L)$ time and memory complexity).

To prevent loops from being present in the paths in the first place, the following procedure is employed:
\begin{enumerate}
  \item Using BFS on the crossing clusters starting from the mid-qubits and constrained to the respective block,
    each site is labeled with the length of the shortest path to the mid-qubit in this block.
    By going in the direction of decreasing length, the shortest path $\mathcal P(s,m(x))$ to a block's $x$ mid-qubit $m(x)$ can be found within its block, starting from any site $s$.
  \item The facing boundaries of all pairs of neighboring blocks $(x_1,x_2)$
    are searched for the pair of sites $(s_1,s_2)$, one in each block, with the least sum of their distances $|\mathcal P(s_1,m_1)|+|\mathcal P(s_2,m_2)|$ from their
    respective mid-qubits $m_1$ and $m_2$, and the bond $(s_1,s_2)$ being open.
  \item With the composite path $\mathcal P(m_1,s_1)\cup(s_1,s_2)\cup\mathcal P(s_2,m_2)$ a loop-free connection between $m_1$ and $m_2$ is found. Although
    paths from the mid-qubits to different neighbors might have sub-paths in common, there will be no loops inside a block due to starting always with the same
    site and the same algorithm for all paths to the boundaries in a given block.

\end{enumerate}

\subsubsection{Reduction to a renormalized lattice}\label{section:reduction}
Instead of the whole square lattice which would require to identify cross-like junctions within the blocks, the procedure will renormalize to a hexagonal
lattice where only T-junctions are required. Of course it will be embedded in the square lattice geometry in the obvious way.
A square lattice would involve crosses the construction of which is not clear when T-junctions have been found and local measurements are used for reduction.
There is, however, an easy way to turn the whole lattice into a square one by using local 
measurements afterwards \cite{NMDB06}.

The procedure to cut out parts of the cluster, or yanking paths straight involves single qubit measurements of the Pauli operators $\sigma_y$ and $\sigma_z$.
Let us briefly summarize the effective action of these operations when applied to a qubit constituting a vertex $a\in V$ of a graph state $|\psi_G\rangle$ with $G=(V,E)$
\cite{HEB04}. The neighborhood of a vertex $a$ the will be denoted by $N_a=\{b:{\text{dist}}(a,b)=1\}$, the subgraph of G induced by $A\subset V$ is $G[A]=(A,\{(a,b)\in E:a,b\in A\})$,
and the complement of a graph $G$ with respect to the set of all possible edges by $G^c=(V,\{(a,b):a,b\in V\}\setminus E)$. The measurement rules now read
\begin{itemize}
  \item $\sigma_y$ measurements perform local complementations: $G'=(G[V\setminus\{a\}]\setminus G[N_a])\cup G[N_a]^c$. So, it swaps all possible edges in its immediate neighborhood.
  \item A $\sigma_z$ ``cuts'' the vertex $a$. The graph representing the resulting state is $G'=G[V\setminus\{a\}]$.
\end{itemize}
The resulting graphs $G'$ actually are only obtained after applying some local unitaries to the neighbors of $a$, which depend on the measurement outcomes, to the remaining qubits.
However, one can also store the effective unitary to apply to each remaining qubit (that is a memory requirement of $O(1)$ per qubit) and adjust the measurement
basis of the subsequent measurements. Because each qubit will be measured out by the end of the computation, it is sufficient to use the second approach to the compensation
of random measurement results.

The first application is to isolate the paths and eliminate the spare sites and dangling ends towards the mid-qubits.
This is achieved by cutting out all unneeded sites by measuring their qubits in the $\sigma_z$ basis.

\begin{figure}[t]
  \includegraphics[width=.5\textwidth]{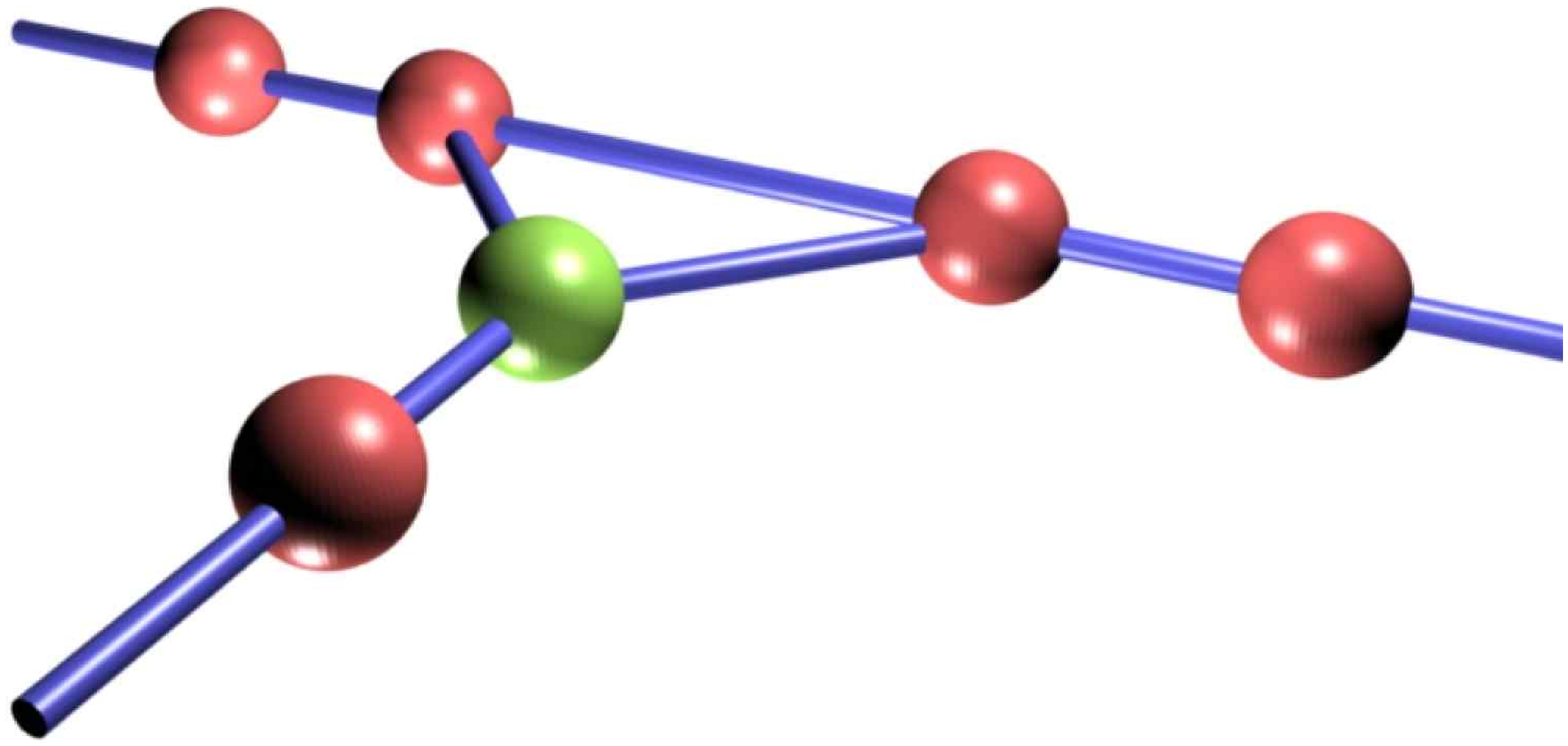}%
  \includegraphics[width=.5\textwidth]{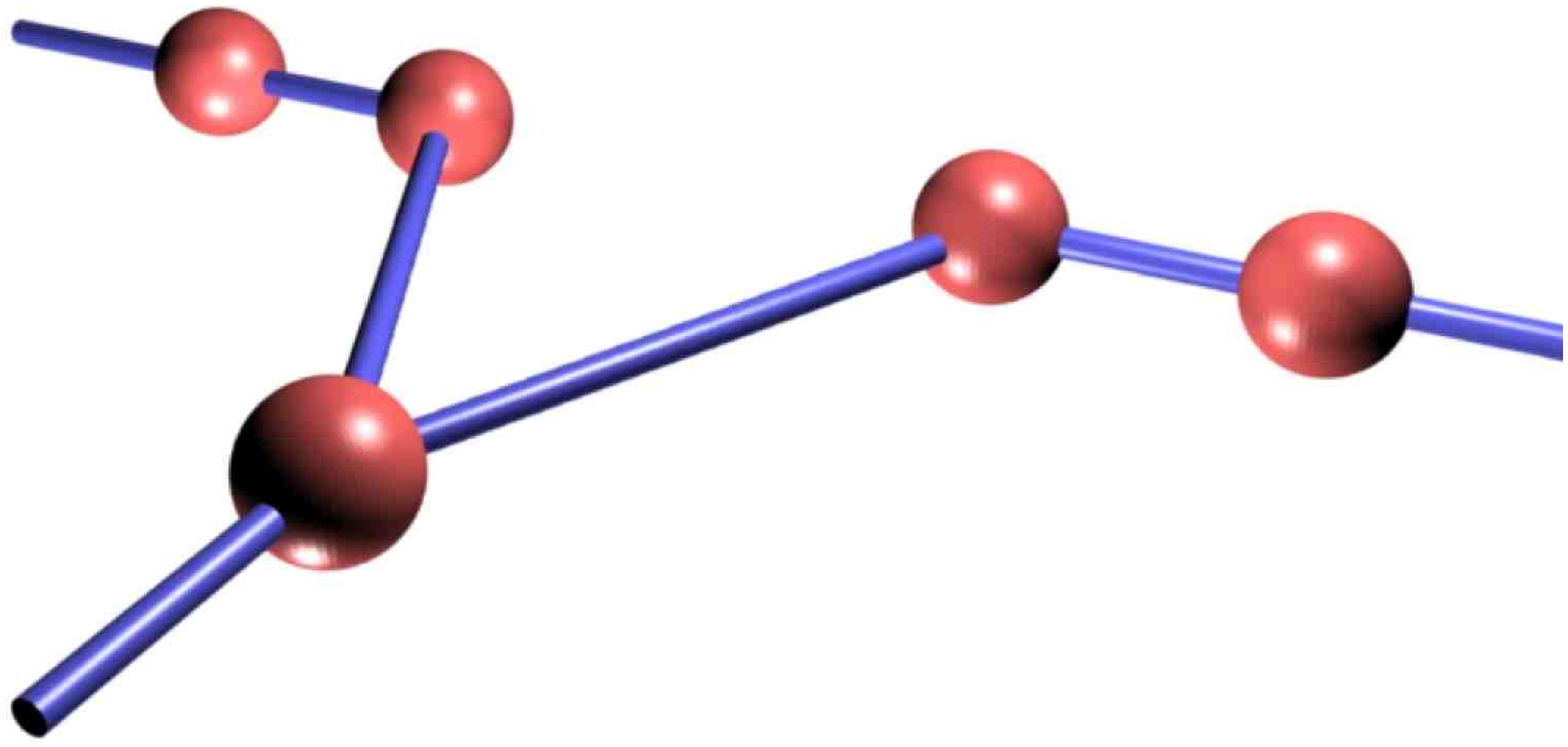}
  \caption{ Effect of a $\sigma_y$ measurement on a triangular junction of a cluster. If the three arms are, as shown, not immediately connected outside the triangle,
    a $\sigma_y$ measurement on the \bwtext{dark}{green} qubit has the effect of destroying the triangle in favor of a T-shaped junction. Due to symmetry,
    any of the three qubits in the triangle could have been chosen. \label{fig:triangles} }
\end{figure}

Now, one is left with a hexagonal lattice where each edge possibly consists of a long path and each site might consist of a triangular structure in the worst case
(depending on the type of lattice used, also the wanted single-qubit sites are possible).
The triangles can be destroyed by suitable $\sigma_y$-measurements, as shown in Fig.~\ref{fig:triangles}. After that, only single-qubit junctions are left,
the paths between which can be shrinked by subsequent application of $\sigma_y$-measurements to a single edge.

Summing up all these contributions we realize that both -- the amount of classical memory and the number of time-steps -- are bounded by $R(L)$ as well.

It should be pointed out that one obstactle in one-way quantum computation is to keep the whole
state in memory. Having fixed the algorithm in advance, the required state size is known
and therefore also the block size for a fixed allowed error rate. Therefore, to grow individual blocks and reduce them to single qubits,
only the neighboring blocks have to exist. This especially allows for growing of the reduced lattice in the time
direction while the computation moves on, requiring only $O(L^{1+\mu})$ qubits to be kept in memory.

\subsubsection{Decreasing the vertex degree}
\paragraph{Choosing the appropriate lattice}

To actually utilize the protocols based on percolation theory the initial resources (i.e., the stars sitting on the lattice sites) should be as small as possible.
As it is more difficult to prepare larger states (this is the problem to be solved in the first place), the lattice with the lowest vertex degree for which
$p_{\rm success}>p^{(c)}$ is still fulfilled will be favorable.

Again, results known in percolation theory, but also specifics of the physical implementation, 
can be used to decrease the vertex degree of the initial states.
First, using star resources, one would have to look for the lattice with the smallest vertex degree that is still suitable for the bond probabilities at hand.
It is not necessary to stay in two dimensions, as the blocking procedure can use high dimensional lattices and renormalize them to two-dimensional square lattices.

\begin{figure}[t]
  \center\includegraphics{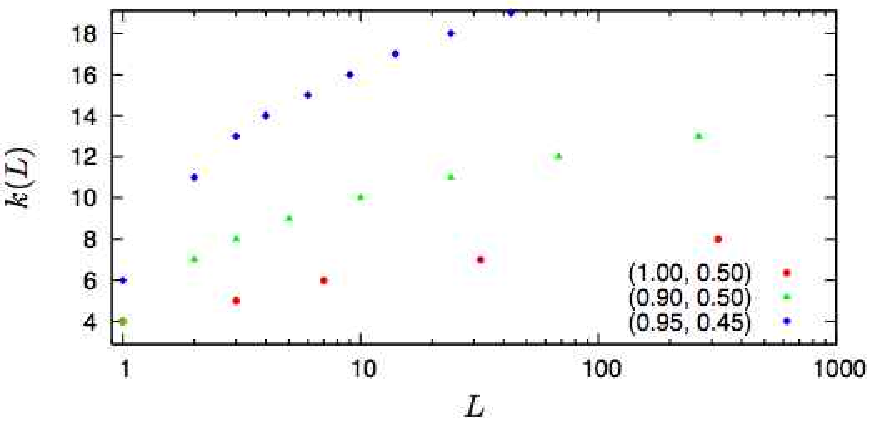}
  \caption{
    Results of Monte Carlo simulations to determine the scaling behavior of the renormalization procedure on the diamond lattice. The dependence of the diamond lattice's block
    size $k\times k\times k$ on the size $L\times L$ of the renormalized square lattice is shown for three different sets of site- and bond probabilities $(p_{\rm s},p_{\rm b})$.
    The threshold for the probability of $\mathfrak U$ occuring was chosen to be $1/2$. $10^5$ blocks of each size were created and used to randomly populate each lattice
    size $10^3$ times.
    \label{fig:diamond_thresholds}
  }
\end{figure}

For example, in the case mentioned above, $p_{\rm success}=1/2$, the smallest vertex degree compatible with $p$ is four, realized by the diamond lattice with
$p_{\diamond}^{(c)}\approx0.389$.
That translates into five-qubit initial states. To see, that the procedure still works for lattices different from qubic, see the results of numerical simulations
on the diamond lattice in Fig.~\ref{fig:diamond_thresholds}.

\paragraph{Covering lattice}

If the entangling operations at hand have the property that one of the qubits survives (like the probabilistic 
parity check ``fusion'' \cite{BR04} in linear optics),
the following property of bond-percolation can further reduce the size of the initial pieces. So far, the sites of the lattice are occupied by single qubits,
the bonds are given by edges in the underlying entanglement graph. If one qubit is left by the entanglement operation we will not think of this one as being a site itself,
but rather belonging to the bond between its neighboring site.

\begin{figure}[t]
  \center
  \includegraphics[width=.26\textwidth]{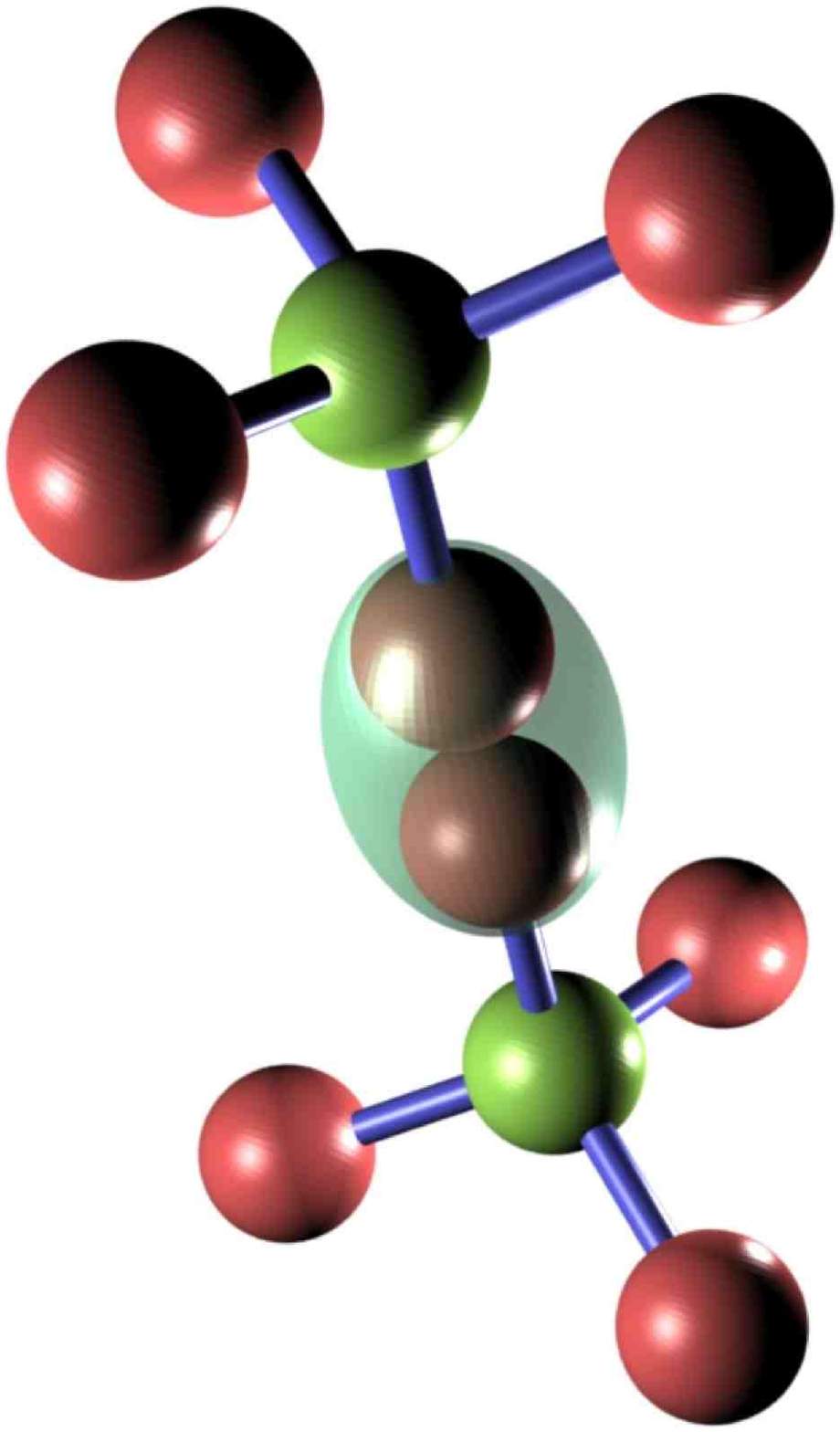}\hspace{.15\textwidth}%
  \includegraphics[width=.26\textwidth]{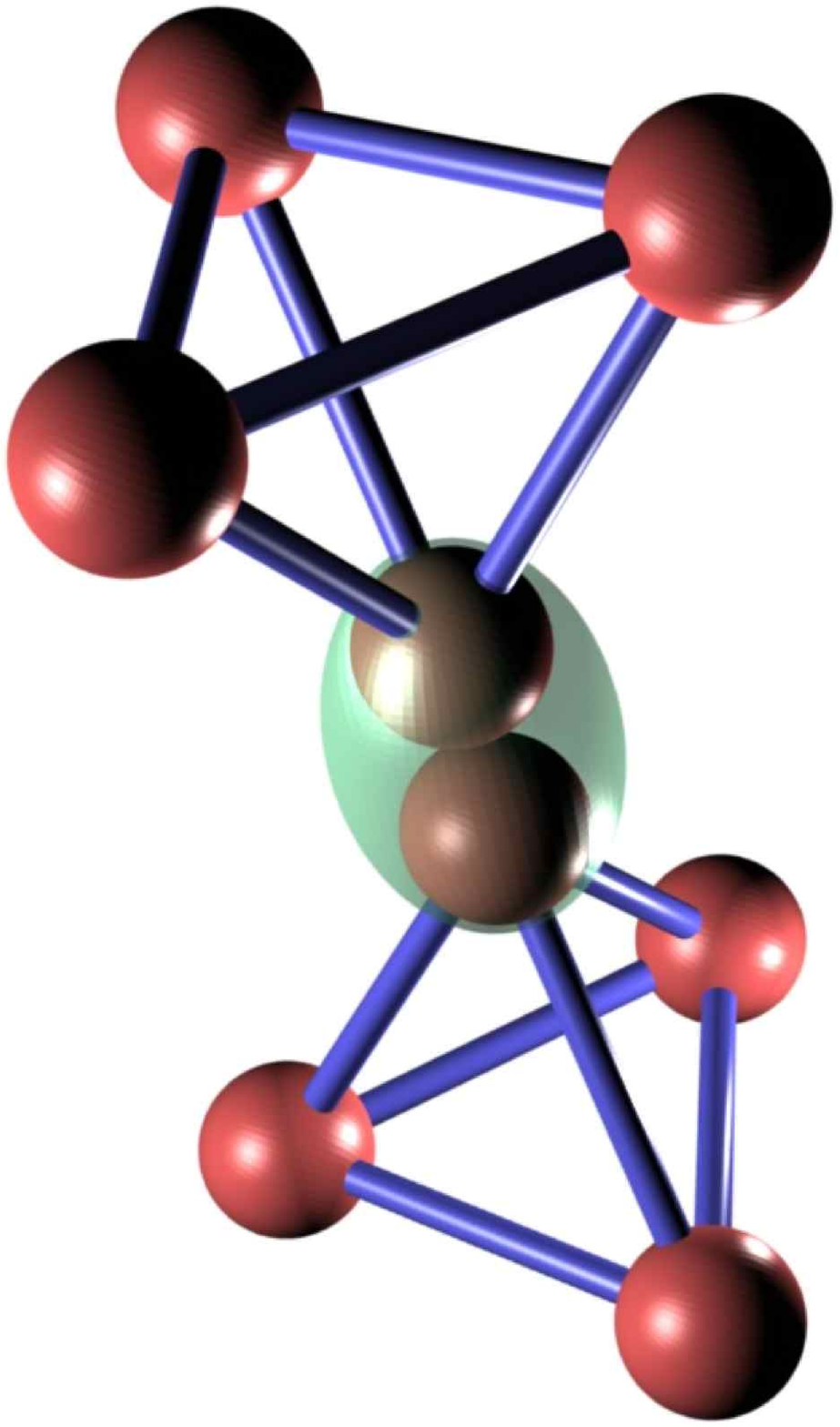}%
  \caption{\label{fig:diamond_pyrochlore}
    Pairs of initial resources for the diamond lattice and its covering one, the pyrochlore lattice. A probabilistic entangling gate is applied
    between pairs of qubits of all neighboring initial states. The central qubit\bwtext{ (dark)}{, here shown in green,} is measured out in the $\sigma_y$-basis,
    resulting in the transformation from diamond to pyrochlore lattice.
  }
\end{figure}

Having such a graph state (which exact lattice type does not matter), we can measure the central qubit $a$ of the initial stars (the \bwtext{dark}{green} one in Fig.~\ref{fig:diamond_pyrochlore})
in the $\sigma_y$-basis. 
%
Given the specific 
structure we have at hand, 
this operation actually performs a transformation from the lattice type we had before to its covering lattice:
now think of the qubits that were sitting on the bonds as proper sites. The old sites have disappeared (they have been
measured out) and the new ones are connected to all the new sites
that were in the same neighborhood of an old site (local complementation).

Intuitively, the covering lattice has the same connectivity properties as the original lattice before. Paths through a star between two arms are existent iff
the star was present and the two entangling operations involving these arms were successful. The same holds on the covering lattice. This property is reflected
by the equation $p_{G,\rm b}^{(c)} = p_{G_{\rm c},\rm s}^{(c)}$, so bond percolation on the original lattice induces site percolation on the covering one.

As the local complementation inside the stars commute with the entanglement operation between them, the central qubit might be neglected from the very start (see Fig.~\ref{fig:diamond_pyrochlore}).
So, one further
qubit can be saved by starting with the fully connected graph state (locally equivalent to the GHZ state) that has one qubit less than the corresponding star. In case
of the diamond lattice, with four-qubit GHZ states (tetrahedral states), the pyrochlore lattice can be built.

\paragraph{Further methods to reduce the conditional 
dependence in linear optical quantum computing}

Surely, if the scheme requires quite large initial stars, they can be prepared with the same tools probabilistically starting from smaller stars.
For a fair assessment, however, the constraints that led to the percolation scheme in the first place have to be imposed here as well.
That is, the restriction to a static setup. Whether composite stars can be used in a static layout now depends
on how the entangling gates work in detail, i.e., what the failure outcomes are. That is of interest due to the fact that in general
a failure in the star preparation step would require back-up steps that do not allow for subsequent application of further entangling gates without re-routing.
A type of gates with suitable properties is the one that acts as a $\sigma_z$-measurement on both qubits on failure.
For example, parity check based gates in linear optics offer this feature.

\begin{figure}[t]
  \psfrag{s}[bc][bc]{success}
  \psfrag{f}[bc][bc]{failure}
  \psfrag{d}[tc][tc]{deterministic\hspace{1.5em}}
  \center\includegraphics[width=.8\textwidth]{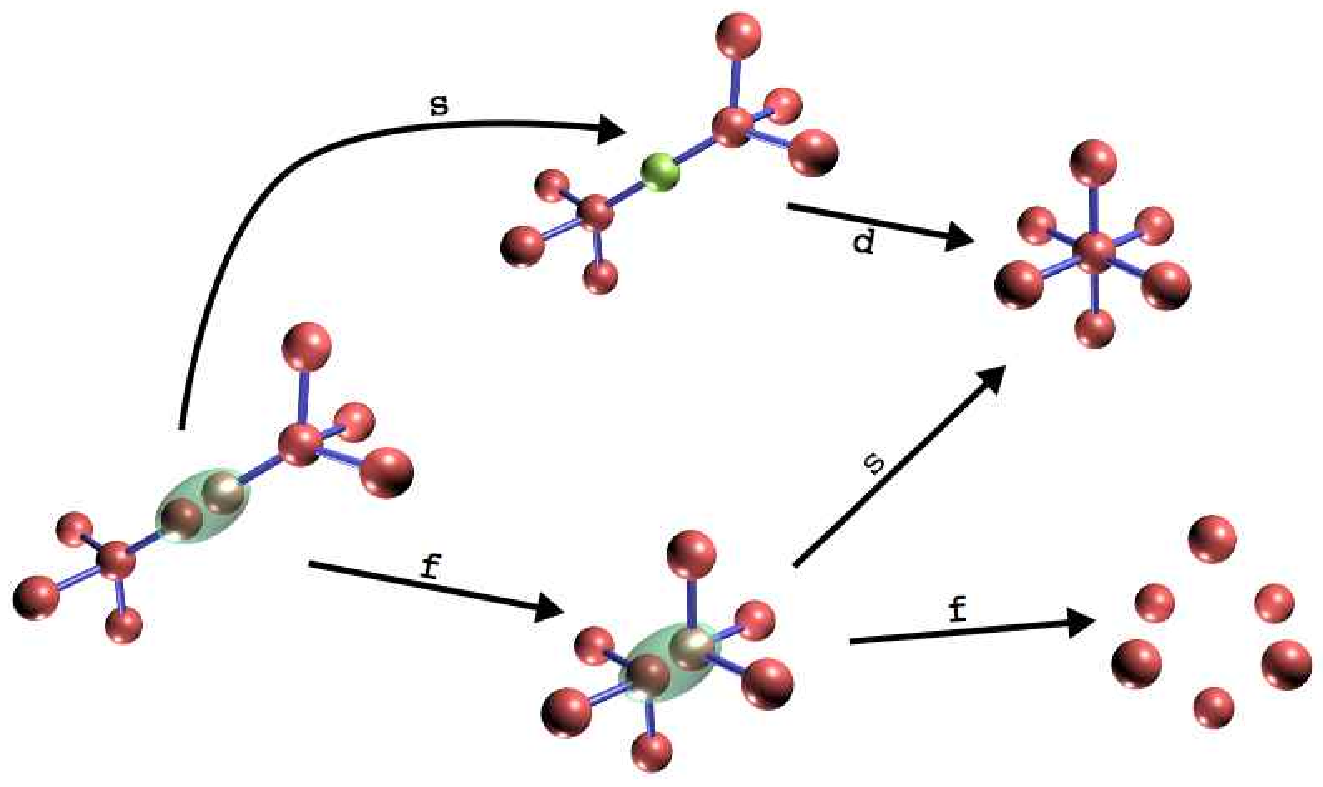}
  \caption{A pair of five-qubit stars (stars with four arms and one central qubit) can be used to create a single seven-qubit star with
    a probability of $p_{\rm success}=1-(1-p_{\rm gate})^2$. With $p_{\rm failure}=(1-p_{\rm gate})^2$ the six arm qubits are separated from each other.
    \label{fig:5fusion} }
\end{figure}

That this feature might actually bring some benefit is shown by the following example (see Fig.~\ref{fig:5fusion}).
Two instances of such an entangling gate are applied to a pair of five-qubit stars, one to a pair of arms and one to the central qubits.
On success of one the ``arm'' fusion gate, the two stars are connected by a two-edge chain. The middle qubit of this chain (\bwtext{dark}{green}) will be
measured out in the $\sigma_x$-basis, leaving the two centers merged in a redundantly encoded qubit, being the center of a star with six arms.
A second application of such a gate on the two qubits of the new center will always succeed due to the entanglement already existing between them.

If, however, the first gate operation failed, the two arms the gate was acting on will be cut of as a consequence of the $\sigma_z$ failure outcome.
The $\sigma_y$ measurement now acts on one part of a product state, leaving the other part -- the two stars -- unchanged.
Now, a second try is possible by application of the entangling operation to the center qubits. The success outcome will be the same as above, on failure
the centers will be cut off, leaving the six qubits in the product 
state vector $|+\rangle^{\otimes 6}$. The failure outcome 
is the result of two consecutive
failures of these entangling gates, so $p_{\rm success}=1-(1-p_{\rm gate})^2$, which is $p_{\rm success}=3/4$ in the case of the linear optics gate
mentioned earlier.

All these operations do not need any classical post-processing or re-routing, therefore, this scheme is suitable to be used in the procedures introduced above.
Because the center qubits are simply cut off when a failure occurs, this procedure and the bond percolation involving the arms are completely independent.
So, both processes together can be modelled as a mixed site/bond percolation with the site probability being $p_{\rm s}=p_{\rm success}$.

As long as the bond probability and the initial state preparation (site-) probability are above the
percolation threshold of the respective lattice, this scheme might be useful to further reduce the size of the
initial star shaped quantum states.

\subsection{Phase transitions in the computational power}
\label{section:transition}
A related yet complementary motivation to look 
at percolation theory in the context of quantum computing is
the question whether a universal resource state can be 
prepared in principle. Surely, above the percolation
threshold a universal resource for quantum computing
can be extracted from a percolating lattice in an efficient
manner, as has been described in the previous section. This
approach, in turn, will fail almost surely below the threshold. 
Indeed, it has been shown in Ref.~\cite{BEF+07} that
for $p<p^{\rm(c)}$ the statistics of all possible measurements (so all possible quantum computations) on the state represented by the resulting 
graph can almost certainly
be efficiently simulated on a classical computer. This 
effect can indeed be seen as a phase
transition in computational power of the resource state 
described by the lattice (caused by the phase transition of the mean cluster size).
Assuming that BQP$\ne$BPP\footnote{This means
that quantum computers are actually more powerful than
classical computers. Although this feature 
is assumed to be highly plausible, the strict separation 
of the complexity classes remains unproven today, not 
unlike the situation for P$\ne$NP.}, 
this phase transition distinguishes between states that can
be used for universal quantum computing and those 
which cannot (see also Ref.~\cite{GESP07}).

This phase transition can be made precise under
the assumption that a certain quantity -- the entanglement width -- provides
a handle on the computational universality of a family of states,
by using it as an order parameter and investigating its properties in the sub- and
supercritical regime.
The entanglement width $E_{\textrm wd}$ is a measure of entanglement, introduced in
Ref.~\cite{NMDB06}. It can be used to characterize the amount of entanglement -- somehow the degree 
of ``non-classicality'' -- contained in a quantum 
system. We refer to Ref.~\cite{NMDB06} for its
definition and properties.
For $d=2$ it can be seen that the amount of entanglement 
contained in the states described
by the lattice jumps from a logarithmic to a linear 
scaling in $L$ at the percolation threshold:

\subsubsection{Supercritical}
Above the respective percolation threshold of a particular lattice of size $L^{2+\mu}$
with $\mu>0$,
an $L\times L$ square lattice can be extracted efficiently, as shown in Chapter~\ref{section:stateprep}.
In Ref.~\cite{BEF+07} an alternative algorithm 
is proposed which achieves $\mu=0$ for $d=2$.

\begin{theorem}[Supercritical states]
  Let $p>p^{\rm(c)}_{2}$. With probability
  \begin{equation}
 \lim\limits_{L\rightarrow\infty}P_p(L) = 1, \end{equation}
  and an overall resource consumption of $O(L^2)$
  an $L\times L$ cluster -- a resource for universal quantum computing -- can be generated.
\end{theorem}

In this approach, in contrast to the renormalization algorithm described in 
Section~\ref{section:stateprep}, the block size of the 
renormalized lattice is not held fixed \cite{BEF+07}, resulting 
in an improved scaling of the resource consumption in case
of the two-dimensional square lattice.

Again, this approach can be used to grow the cluster state while computation moves on.
With a fixed width new layers can be added on top, allowing for subsequent identification of new horizontal paths
(and their connections to the underlying lattice). So it is not required to keep the whole lattice in memory,
rather only a set of qubits of size $O(L)$.

A natural substructure of a percolating $n\times n$ lattice is 
already provided by the fact
that there exist at least $m=O(n)$ vertex-disjoint crossing 
paths (which shall be event $\mathfrak{G}_m(n)$) in either dimension almost certainly in the limit of large $n$.

Consider the event of having at least one left-to-right crossing path in a two-dimensional $n\times n$ block,
which happens with probability 
\begin{equation}
	P_p(\mathfrak{A}(n))\ge 1-\exp(-g(p)n^2), 
\end{equation}
$g(p)>0$ (Eqn.~\ref{eqn:block}).
When this event is still happening when changing the state $r$ arbitrary edges on the lattice,
there have to be at least $r+1$ edge-disjoint left-to-right crossings in this block.
With $r=\beta(p)n$ an inequality from Ref.~\cite{Grimmett} allows us
to write
\begin{equation}
  P_p(\mathfrak{G}_{\beta(p)n}(n)) \ge 1-\left(\frac{p}{p-p^{(c)}_2}\right)^{\beta(p)n}{\mathrm e}^{-ng(p)} .
\end{equation}
In the limit of $n\rightarrow\infty$ this probability approaches unity, iff
\begin{equation}
  g(p)-\beta(p)\log\left(\frac{p}{p-p_2^{\rm(c)}}\right) > 0,
\end{equation}
which can always be achieved by suitable choice of $\beta$ for a given $p$.
Therefore, in the limit of large $m$, there are almost certainly $m$ edge-disjoint open left-to-right crossing paths in
a square lattice of side length $n=m/\beta(p)$. In other words, the number of these crossings scales as $O(n)$.

\paragraph{Algorithm}
\begin{figure}[t]
  \includegraphics[width=.5\textwidth]{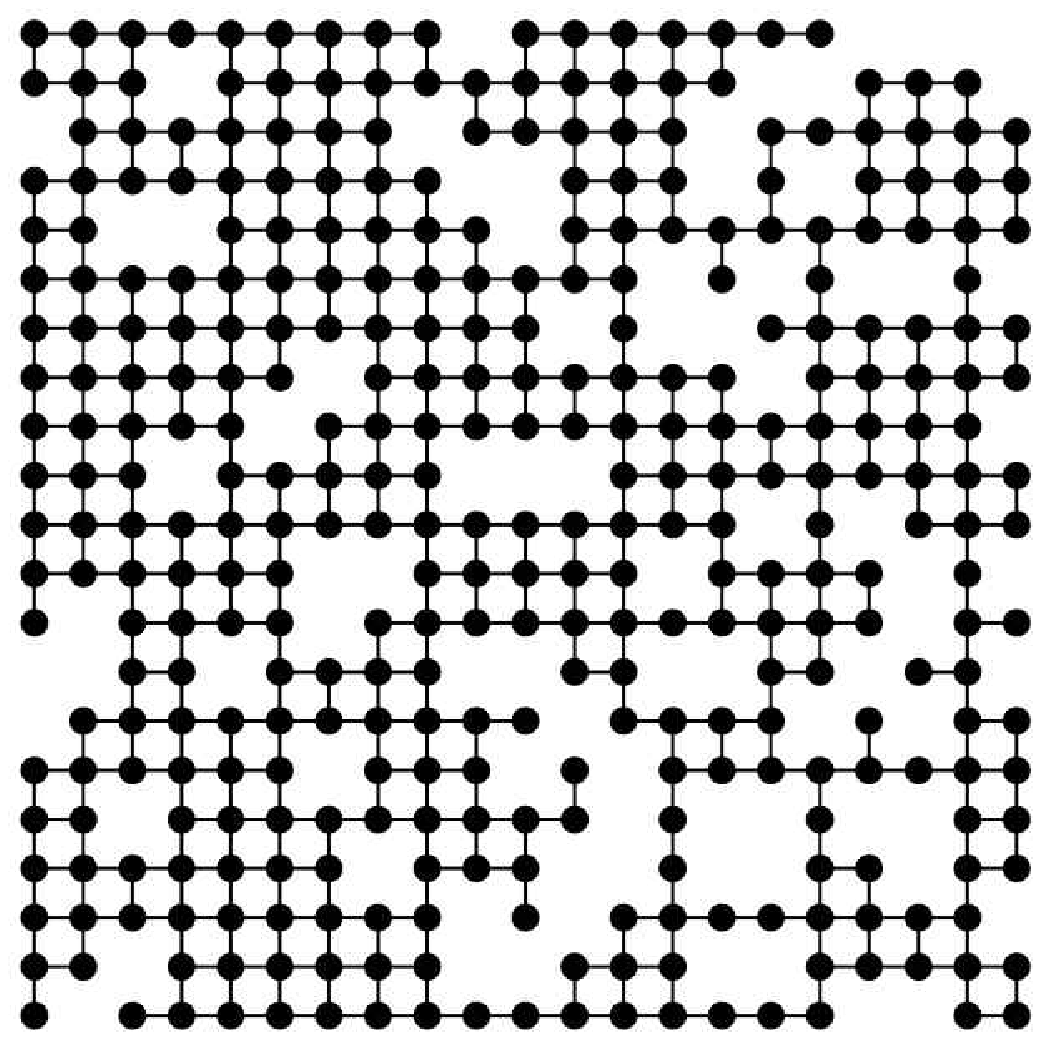}%
  \includegraphics[width=.5\textwidth]{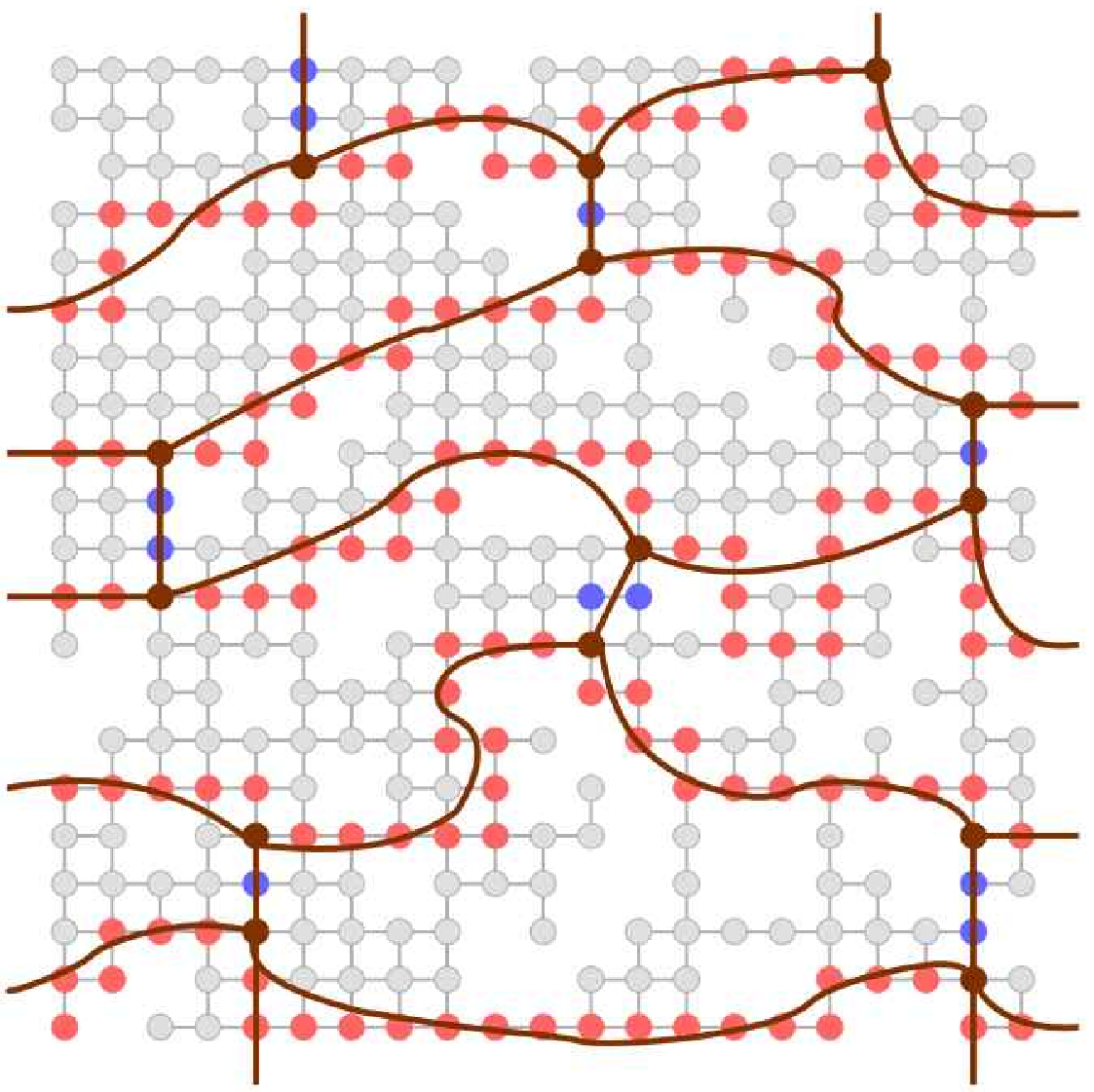}
  \caption{ \label{fig:algorithm}
    Illustration of the algorithm to extract an $L\times L$ lattice with only constant overhead per site. The left picture shows the initial percolated lattice.
    At the right, the identified paths after alternating bridge decomposition and error correction, as well as the final graph after the quantum measurements are shown.}
\end{figure}
The challenge now is to identify $O(L)$ suitable paths such that the reduction to a fully populated regular lattice
with the tools of Chapter~\ref{section:reduction} is possible. 
In Ref.~\cite{BEF+07} a very detailed description of an algorithm achieving this goal can be found,
some of the steps of which are illustrated in Fig.~\ref{fig:algorithm}.
The important stages are the following:

\begin{enumerate}
  \item \emph{Path identification}:
    An algorithm from maze solving, the \emph{right-hand wall follower}, which
    can identifies the right-most path along a wall (which will be
    the surface of the graph, or the previous path), is used.
    By applying it consecutively it finds a maximum set of non-intersecting crossing open paths
    in either direction.

    Vertical paths are found by using every third path
    that result from the right-hand wall follower.
    For horizontal paths, the \emph{2-local} version of this algorithm is used,
    which means that a graph distance on the underlying lattice to the previous path of at 
    least $3$ is ensured.

    These gaps between neighboring paths ensure that the 
    reduction can be performed in the end. A distance of $1$ between neighboring horizontal paths is surely not enough
    when it comes to the shortening of intermediate parts by means of $\sigma_y$ measurements. Having one qubit
    in between turns out to be not enough where intersections with vertical paths occur.

  \item \emph{Alternating bridge decomposition}:
    While fixing paths in the horizontal direction, the vertical ones are cut into segments between neighboring
    horizontal paths (\emph{bridge decomposition}). By only using every other of these segments (alternating bridge decomposition), a hexagonal structure is imposed,
    simplifying the reduction stage as before, while still giving rise to a universal resource state.
    Due to the choice of
    using every third vertical path,
    the graph has essentially the same topology as a hexagonal lattice, up to local errors at the crossings.
    For an in-depth analysis we refer to Ref.\ \cite{BEF+07}.

    The bridge decomposition further assures a simple topology of the crossing points, because multiple crossings and common
    parts of horizontal and vertical paths are cut out.

  \item \emph{Correction of local errors}: All that is left at this stage is a hexagonal structure where the bonds consist of paths the vertices of which have vertex degree $2$,
    and the sites consist of a couple of vertices with degree not necessarily $3$.
    More specifically,
    they contain vertices of degree $\ge3$ on the horizontal paths 
    (the \emph{abutments}) and one vertex per crossing of degree $\ge2$ at
    the beginning (or end, respectively) of a vertical path. By appropriately 
    removing the vertices in the middle of the abutments from the graph, the crossings are reduced to
    single vertices of degree $3$ (a T-shaped junction), thus allowing for 
    reduction of the lattice by local Pauli measurements.
\end{enumerate}
After this classical identification process is done, the graph is reduced to a hexagonal lattice using local measurements as in Section~\ref{section:reduction}.

\paragraph{Entanglement width}
An $L\times L$ cluster state has an entanglement width of $E_{\textrm wd} = O(L)$. Due to the fact
that the entanglement width does not increase during one-way computation (which also is what is
effectively done to extract the cluster from the percolated lattice in all known proposals, see also
Chapter~\ref{section:stateprep}), the 
extracted cluster can be used to derive a lower bound to this
entanglement width,
\begin{equation}
 E_{\textrm wd}(p > p^{\rm(c)}) \ge O(L) . 
 \end{equation}
This bound holds for any regular lattice with $d\ge 2$ above its respective percolation threshold.

\subsubsection{Subcritical}
Below the percolation threshold the following behavior 
is observed:
\begin{theorem}[Subcritical states]
  For $p<p^{\rm(c)}_{2}$, the corresponding quantum states can be simulated efficiently on a classical computer.
\end{theorem}
The idea behing the proof is that the
entanglement width is bounded by the entanglement width of the
largest connected component \cite{BEF+07}. Below the percolation threshold, 
this is of size $O(d \log L)$ \cite{Bazant07}, therefore bounding
\begin{equation}
 E_{\textrm wd} \le O(\log L) .\end{equation}
Making use of Ref.~\cite{Bazant07}, it is shown that 
this bound holds for $2\le d\le6$.
Further, the logarithmic scaling of the size of the largest connected component results in the computational 
space to be of size ${\rm poly}(L)$, so efficiently simulatable by classical means.

\section{Quantum repeater networks for quantum communication}
\label{section:networks}
\subsection{General concepts}

Another quite natural application of percolation ideas in the context of 
quantum information processing is to find ways of designing 
quantum networks for quantum communication. This is now an 
application no longer related to quantum computation, but to quantum 
communication and quantum key distribution. Here, the concept
of a network emerges quite naturally, and it seems meaningful
to ask to what extent percolation ideas can help here. 

The setting we will discuss in this section -- reviewing the content of
Refs.\  \cite{ACL07,PCAL07} -- is slightly different from the previous ones:
We will no longer make strict use of classical percolation ideas to 
identify threshold behavior in the possibility of preparing some quantum states. Instead,
but in a closely related fashion, we will ask to what extent
one can modify quantum measurement schemes to realize percolation-like settings which outperform standard bond 
percolation when naive measurement patterns are used. It will hence be not be sufficient to look at classical properties of 
quantum states as before, but we will have to consider the quantum states as such.

The idea of {\it quantum communication} or, more specifically, of 
{\it quantum key distribution} 
plays a seminal role in quantum information science: 
Indeed, whenever two laboratories share quantum systems in 
highly entangled states, by appropriate measurements and classical postprocessing,
a {\it secret key} can be extracted from the measurement data. 
This key can then be used for the secure transmission of 
classical data. For a review on this idea, which has led to numerous
experimental implementations, and based on which commercial devices are available, see, e.g., Ref.\ \cite{GRTZ02}. 
For ``sufficiently entangled'' states
one can indeed show that the resulting key distribution is unconditionally
secure, in the sense that 
the security proof does not rely on unproven assumptions on the 
hardness of certain mathematical problems. Instead, it is merely assumed
that an eventual eavesdropper could have access to any technological device,
but is constraint by the laws of quantum theory as such in his or her possible 
action. 

The functioning of such ideas, however, relies on the availability
of sufficiently pure and entangled quantum states. If -- in a sense -- 
too much noise is in the system, which is the situation encountered
in distribution of entangled state through noisy channels over large distances,
then the key will no longer necessarily be secure. In such a
situation, ideas of 
{\it quantum state distillation} or of {\it quantum repeaters}
offer a solution: This amounts to a fragmentation of the whole transmission
line to a sequence of {\it repeater stations}, or, more generally, to a network of 
repeater stations.  And here, needless to say, is where percolation ideas come in.
 
We will start by describing the setting of Ref.\ \cite{ACL07}: 
Consider some lattice, the vertices of which are identified with repeater stations, or
laboratories in which one can perform quantum operations. Between such 
vertices, non-maximally entangled states can be created. Once two vertices
share entangled  quantum systems, we will say that they are connected
by a  {\it ``quantum edge''}. Following this entanglement distribution process, 
only ``local'' operations constrained to the qubits of a repeater station are
possible, coordinated by classical communication. This is the setting
of {\it local operations with classical communication} (LOCC), which is the
standard paradigm when considering distributed quantum systems. Based
on such operations, purification protocols \cite{BVK99,Vidal99} can 
transform a single pair or a chain of two 
pairs into a maximally entangled
pair, the equivalent to an open bond. This process is intrinsically probabilistic,
which represents the bond probability. The probability of success of such a protocol (and with it the bond probability)
depends not only on the amount of entanglement that is available in the very beginning,
but also on the distillation strategy employed. Therefore, we cannot assign unique
bond probabilities to a lattice with quantum bonds.

\subsection{Classical entanglement percolation}

We will now compare two different settings,
following Refs.\ \cite{ACL07,PCAL07}. The first approach is to 
make use of natural physical measurements where each of 
the quantum edges is purified separately. This then leads to a 
familiar edge percolation problem: A quantum edge corresponds 
to sharing a non-maximally entangled state with a state vector of 
the form 
\begin{equation}
	|\varphi\rangle={\lambda_1}^{1/2}|0,0\rangle+
	{\lambda_2}^{1/2}
	|1,1\rangle .
\end{equation}
 $\lambda_1$ and $\lambda_2$ are the {\it Schmidt
coefficients} satisfying $\lambda_1+\lambda_2=1$, assuming 
that $\lambda_1\ge\lambda_2$. For simplicity, all 
non-maximally entangled states are taken to be pure.
It is known that there exist local quantum operations
assisted by classical communication (LOCC) 
that transform $|\varphi\rangle$ into a maximally entangled 
pair with state vector
$2^{-1/2}(|0,0\rangle+|1,1\rangle)$, 
with a probability of success
of 
\begin{equation}
	p_{1}=\min(1,2(1-\lambda_1)).
\end{equation}	
This probability of success will be referred to 
as {\it singlet conversion probability}
(SCP) of a single quantum edge. In general, having a LOCC
conversion protocol
means that there exist operators $M_A^{(k)},M_B^{(k)}$ 
satisfying $\sum_{k}(M_A^{(k)})^\dagger M_A^{(k)}\leq {\mathbbm{1}}$
and $\sum_{k} (M_B^{(k)})^\dagger M_B^{(k)}\leq 
{\mathbbm{1}}$ 
which can be implemented with LOCC
such that 
\begin{equation}\label{One}	
	(M_A^{(k)}\otimes M_B^{(k)})
	|\varphi\rangle=(p^{(k)})^{1/2} (|0,0\rangle+|1,1\rangle)2^{-1/2},
\end{equation}	
$p_1=\sum_k p^{(k)}$. In this case, the filtering is quite simple:
we merely 
need to consider a single successful outcome, $k=1$, and
take 
$M_A^{(1)}=(\lambda_2/\lambda_1)^{1/2} 
|0\rangle\langle 0| + |1\rangle\langle 1|$ and $M_B^{(2)}={\mathbbm{1}}$. 
As can be readily
verified, this leads to a maximally entangled state with the above
probability of success. This filtering process is called 
{\it Procrustean method} \cite{BBPS96}. 

Once one has a maximally entangled pair, we encounter the 
situation as before, and we say the corresponding edge is open, 
in the sense of usual edge percolation, and as being used 
throughout this chapter. If the protocol fails, one is led to an unentangled state, and
the edge is closed. Hence, $\lambda_1$ 
governs the percolation behavior and defines the edge probability.
If one is above the percolation threshold
of the respective lattice, a connected path can be identified 
almost certainly between any two vertices $A$ and $B$ of the lattice:
Let $\mathfrak c(A,B)$ denote
 the event that $A$ and $B$ are connected by maximally entangled bonds.
Then, 
\begin{equation}
	\lim_{{\text{dist}}(A,B)\rightarrow\infty}P(\mathfrak c(A,B))>0
\end{equation}
if and only if $p_1>p^{(c)}$. 
Physically, this means that $A$ and $B$ share a perfect quantum
channel through which the state of a single qubit can be sent in a lossless
fashion. This setting which is the natural analogue of edge 
percolation will be called {\it classical entanglement percolation (CEP)}.

\subsection{Quantum percolation strategies}
\label{section:qperc}
The quantum character of the involved states (``quantum edges''), however,
allows for some improvement. The main observation is that for quantum systems, 
one does not necessarily have to perform
the above measurements, leading individually to open or closed edges, but
can resort to appropriate local collective operations. If the task is to 
achieve a perfect quantum channel between any two vertices $A$ and $B$ in a
lattice, then we have achieved this when having a maximally entangled
pair of qubits between $A$ and $B$ at our disposal. This process should
succeed in the best possible fashion. The aim hence 
is to maximize the SCP, so the probability to achieve a maximally 
entangled pair between two vertices $A$ and $B$,
but not necessarily exploiting the above CEP.  

To exemplify the mechanism, let us first consider a one-dimensional
chain with three vertices, $A$, $B$, and $C$, such that $A$ and $B$, and
$B$ and $C$ share a state with state vector $|\varphi\rangle$ each 
(they
are connected by a quantum edge). This is a quantum repeater
situation. Then, clearly, applying 
the above filtering corresponding to CEP twice,
one succeeds with a probability $p_{1}^2$.

But, since vertex $B$ holds two quantum systems, one can also do a 
collective local operation. The SCP 
asks for the optimal probability
$p=\sum_k p^{(k)}$ 
such that
\begin{equation}\label{OptimRep}
	(M_A^{(k)}\otimes M_{B_1,B_2}^{(k)}\otimes M_C^{(k)})
	|\varphi\rangle_{A,B_1} |\varphi\rangle_{B_2,C}=
	(p^{(k)}/2)^{1/2}(|0,0\rangle_{A,C}+|1,1\rangle_{A,C})|0,0\rangle_{B_1,B_2}.
\end{equation}	
Clearly, $p\geq p_{1}^2$, as the protocol in Eq.\ (\ref{OptimRep})
includes the prescription of CEP where one tries to purify the
entanglement between $A$ and $B_1$ and $B_2$ and $C$ individually.

We can also easily obtain an upper bound to the probability 
of success: The class of protocols defined 
in Eq.\ (\ref{OptimRep}) 
is included in the one of Eq.\ (\ref{One}) in the case when
$A$, $B_1$, and $B_2$ are treated as a single system. Doing such 
collective operations (instead of only in $B_1$ and $B_2$)
can only improve the probability of success, hence this
upper bound. 

Somewhat surprisingly, it can be shown that this bound can 
indeed be achieved \cite{BVK99,Vidal99}. 
In $B_1$ and $B_2$ one performs a
joint measurement $k=1,\dots,4 $ with 
\begin{equation}
M_{B_1,B_2}^{(k)}=|\psi^{(k)}
\rangle_{B_1,B_2}\langle  \psi^{(k)}|_{B_1,B_2}, 
\end{equation}
where
\begin{eqnarray}
	|\psi^{(1)}\rangle &=& (|0,0\rangle + |1,1\rangle)2^{-1/2},\,\,
	|\psi^{(2)}\rangle = (|0,0\rangle - |1,1\rangle)2^{-1/2},\\
	|\psi^{(3)}\rangle &=& (|0,1\rangle + |1,0\rangle)2^{-1/2},\,\,
	|\psi^{(4)}\rangle = (|0,1\rangle - |1,0\rangle)2^{-1/2}.
\end{eqnarray}
Two of the outcome probabilities of this four-outcome 
measurement are equal to
$p_{\text{max}}=(\lambda_1^2 + \lambda_2^2)/2$,
whereas the two further outcome probabilities
are given by
$p_{\text{min}}=\lambda_1  \lambda_2 $.
This local filtering is in fact nothing but a collective 
{\it ``Bell''-measurement} in 
the $\sigma_z\sigma_z$-basis. This procedure reaches
the above upper bound, and hence
\begin{equation}
	p= p_{1}.
\end{equation}
This scheme, simple as it is, already shows that with 
collective operations, one can 
improve the involved probabilities, from $p_1^2$ to $p_1$. 
Note that here the SCP is taken as a figure of merit, 
whereas other figures of merit, like the average final 
entanglement quantified in terms of an entanglement measure,
can also be meaningful quantifiers of success of the scheme
\cite{PCAL07}. Also, the same protocol is optimal in case
one has two different pure states in the Schmidt basis at hand.

\subsection{Absence of improvement for one-dimensional systems}

A first natural instance where this process can be applied is the
case of one-dimensional systems. In this case, 
it turns out, there is no improvement possible. The improvement
that is possible for two quantum edges does not persist in the
asymptotic limit of a quantum chain, and an 
exponential decay of the SCP with the distance between
two vertices of the chain cannot
be avoided \cite{ACL07}. Moreover, 
the only way of having unit SCP is to have a chain of
maximally entangled pairs in the first place. The argument
to show this makes use of the entanglement measure of
the {\it concurrence} of a two-qubit state. For a state
vector $|\varphi\rangle = \sum_{j,k} T_{j,k}|j,k\rangle$ its
concurrence is found to be $C(\varphi)= 2|\text{det}(T)|$,
in terms of the $2\times 2$-matrix $T$. In the situation of
having a one-dimensional chain of repeaters 
involving $N+1$ quantum edges and hence $N+w$ 
vertices, involving $N$ correlated 
measurements, one then
finds that the maximum average concurrence of the
first and the last qubit in the chain is given by
\begin{equation}
	C=\sup_{\cal M}
	\sum_{r_1,\dots, r_N}
	2 \left|\text{det}
	\left(
	|\varphi_1\rangle\langle \varphi_1| 
	M^{(r_1)}_1 
	|\varphi_2\rangle\langle \varphi_2| 
	\dots M^{(r_N)}_N 
	|\varphi_{N+1}\rangle\langle\varphi_{N+1}|\right)\right|.
\end{equation}
In this expression, ${\cal M}$ stands for all measurements,
whereas the $M^{(r_i)}_i$ are $2\times 2$ matrices 
specifying the $r_i$-th outcome at site $i$, corresponding
to the state vector $|\mu_i\rangle = \sum_{j,k} (M^{(r_i)}_i)_{j,k}|j,k\rangle$ of this specific measurement outcome. From this
expression demonstrating the exponential decay of the 
concurrence with the distance between two vertices in a 
one-dimensional chain, one can also derive the exponential
decay of the SCP. Hence, while for two repeaters one can find a 
better probability of success, this does not lead to a different
asymptotic performance of this type of quantum percolation
strategy.

For other lattices, different from simple one-dimensional chains, 
one can improve the situation, however, 
compared to classical entanglement percolation, as we will
discuss in the subsequent subsection.
In this way, a lattice having some quantum edges
can be transformed into a different lattice with maximally 
entangled pairs being randomly distributed, whereas
the distribution will depend on the strategy employed (e.g., CEP or collective operations).
This approach can outperform methods 
relying solely on classical probabilities and equivalences between 
these lattices (e.g., covering, matching, or dual lattices). This 
is an effect which is caused by the quantum nature of the 
bonds.

\subsection{Examples of quantum percolation strategies}
\paragraph{Hexagonal lattice}

An example to highlight the difference to a purely classical 
approach is the following: One starts from the situation 
where neighbors on a hexagonal lattice share
two specimens of the same state. This situation is very natural 
in the quantum case: One simply has to distribute a state of the 
same kind twice. The aim is to establish a connected open path 
between arbitrary vertices (taken from a triangular sublattice).

\begin{theorem}[Singlet conversion probability in the hexagonal lattice] \label{thm:hexa}
Consider a hexagonal lattice, where nearest neighbors share $|\varphi\rangle^{\otimes2}$
with $|\varphi\rangle = {\lambda_1}^{1/2}|0,0\rangle+ {\lambda_2}^{1/2}	
|1,1\rangle $. Then, if $\lambda_1> (1/2 + \sin(\pi/18))^{1/2}$ but
$1-\lambda_1 > \sin(\pi/18)$,
CEP does not lead to edge percolation, whereas joint measurements result in an infinite connected
cluster.
\end{theorem}

We start by describing the situation for CEP: Each pair of neighbors shares 
$|\varphi\rangle^{\otimes2}$, so the optimal SCP of transforming these two
copies into a maximally entangled state (an open edge) is
$p_{1}= 2(1-\lambda_1^2)$,
where the $\lambda_1^2$ originates from the fact that we have two specimens
at hand. It is easy to see that with the above choice for $\lambda_1$ one 
arrives at an edge probability of 
\begin{equation}
	2(1-\lambda_1^2)<1- 2 \sin (\pi/18)= p_{\hexagon}^{\rm(c)} \approx 
	0.6527.
\end{equation}
Hence, with this strategy one falls below the percolation threshold of the hexagonal lattice.

\begin{figure}[t]
  \includegraphics[width=.45\textwidth]{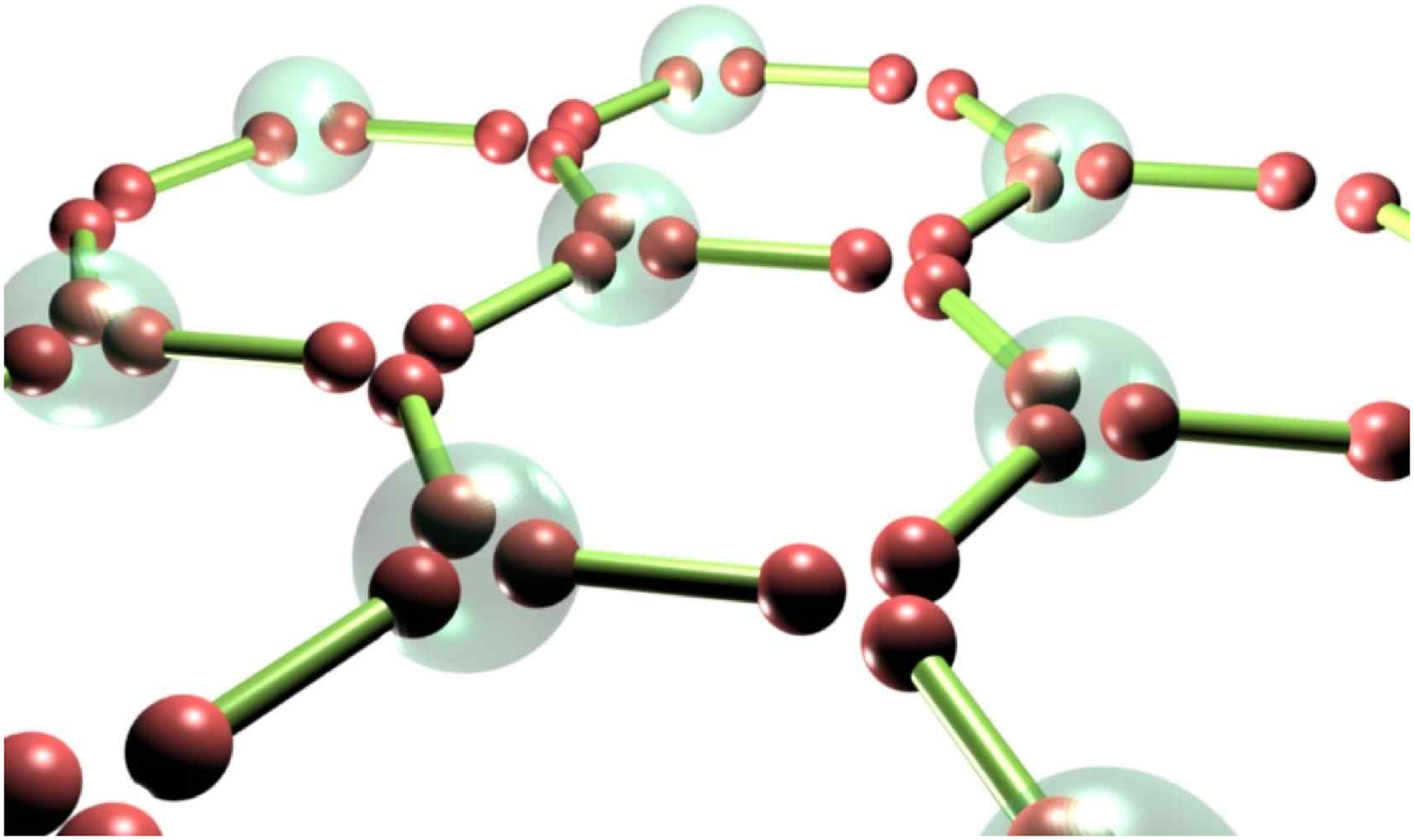}\hfill%
  \includegraphics[width=.45\textwidth]{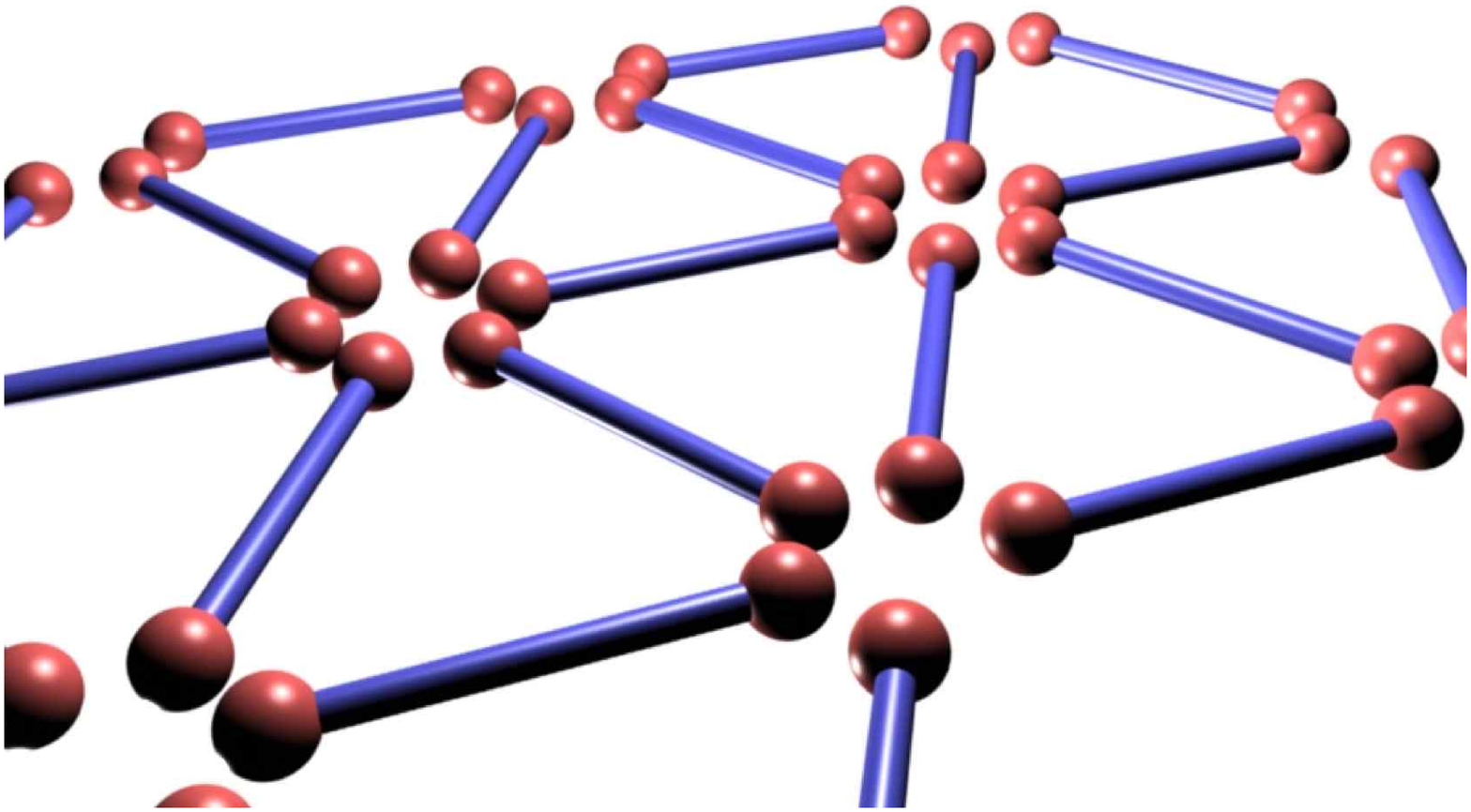}

  \includegraphics[width=.25\textwidth]{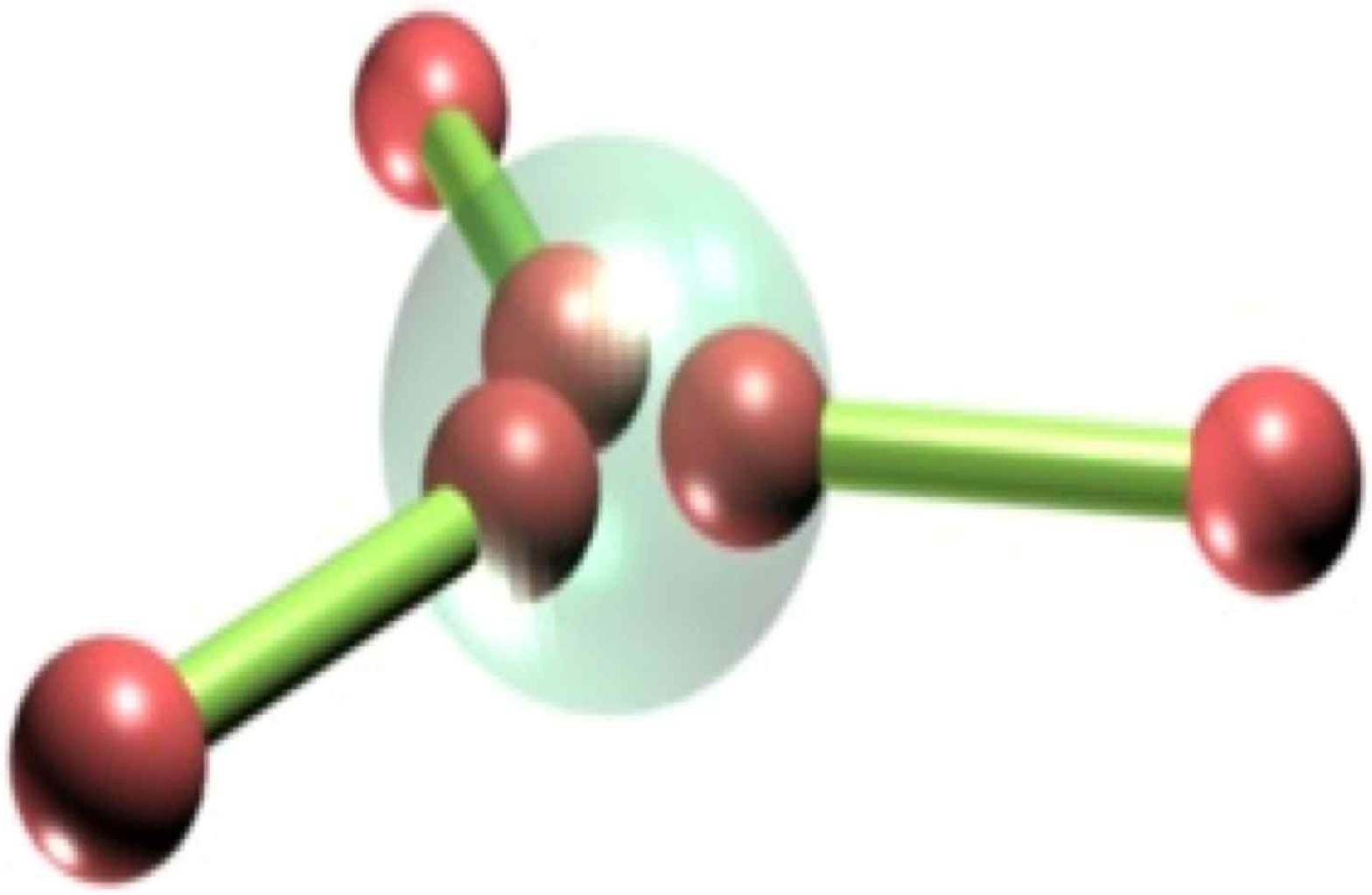}%
  \includegraphics[width=.25\textwidth]{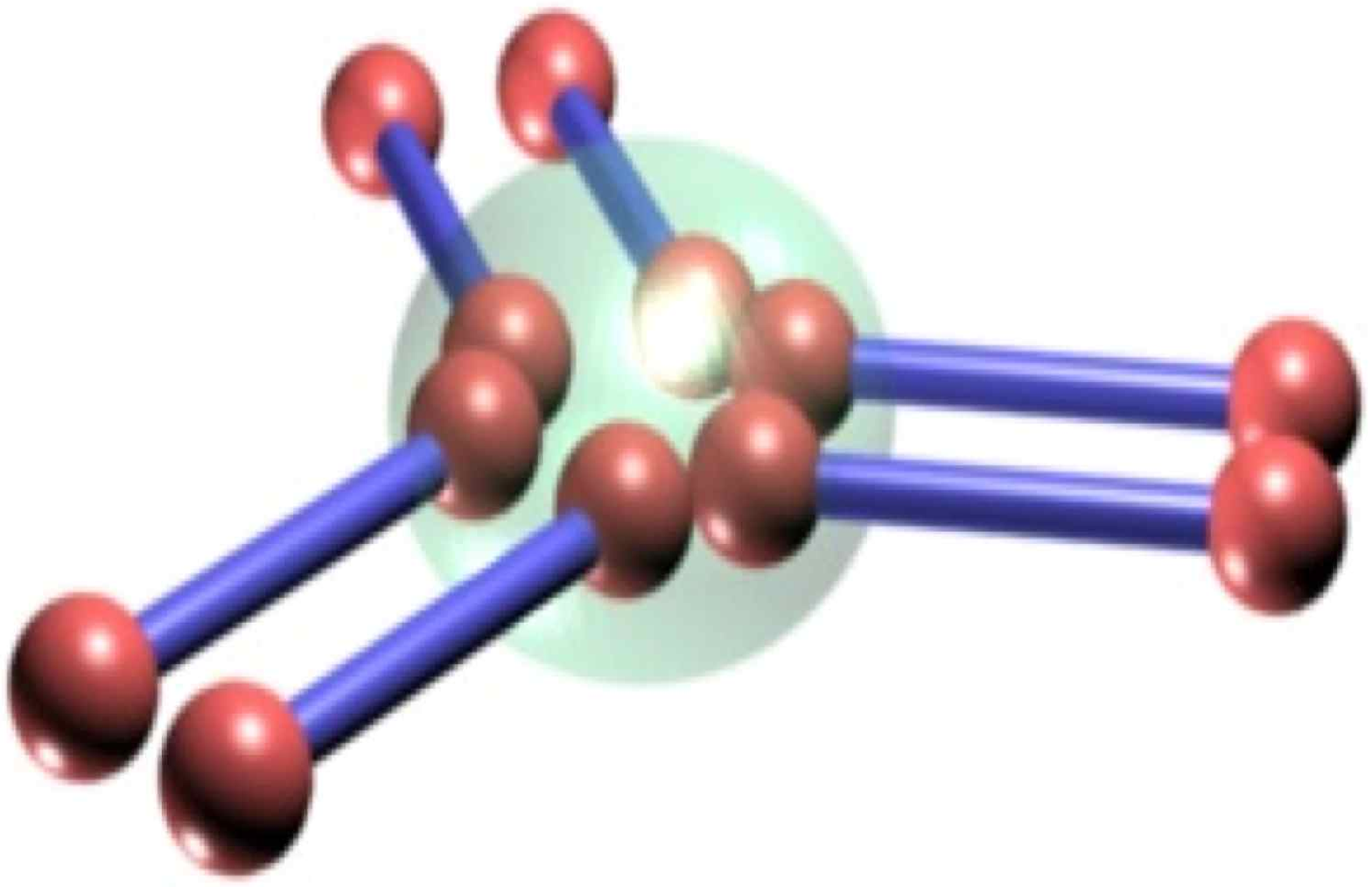}%
  \includegraphics[width=.25\textwidth]{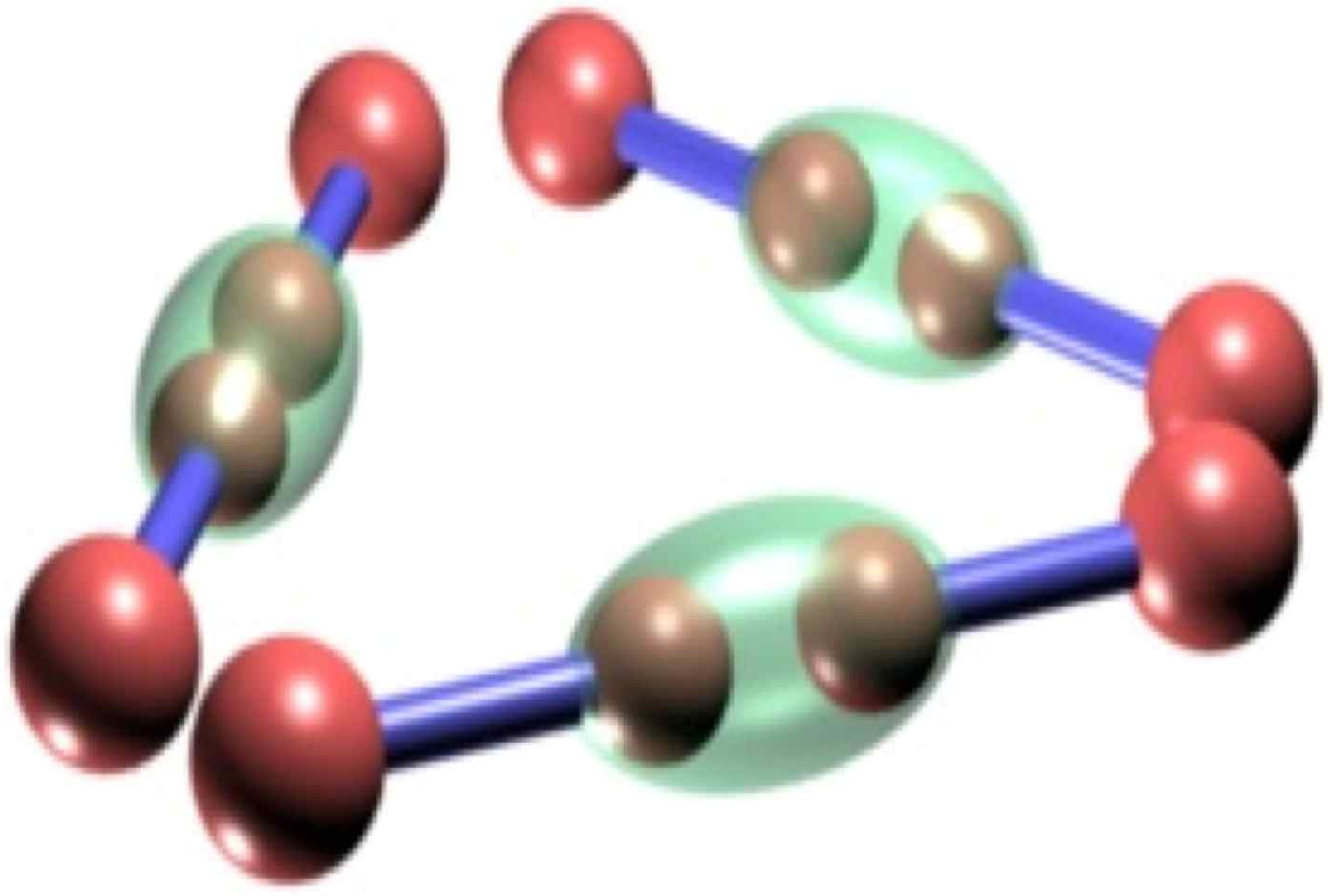}%
  \includegraphics[width=.25\textwidth]{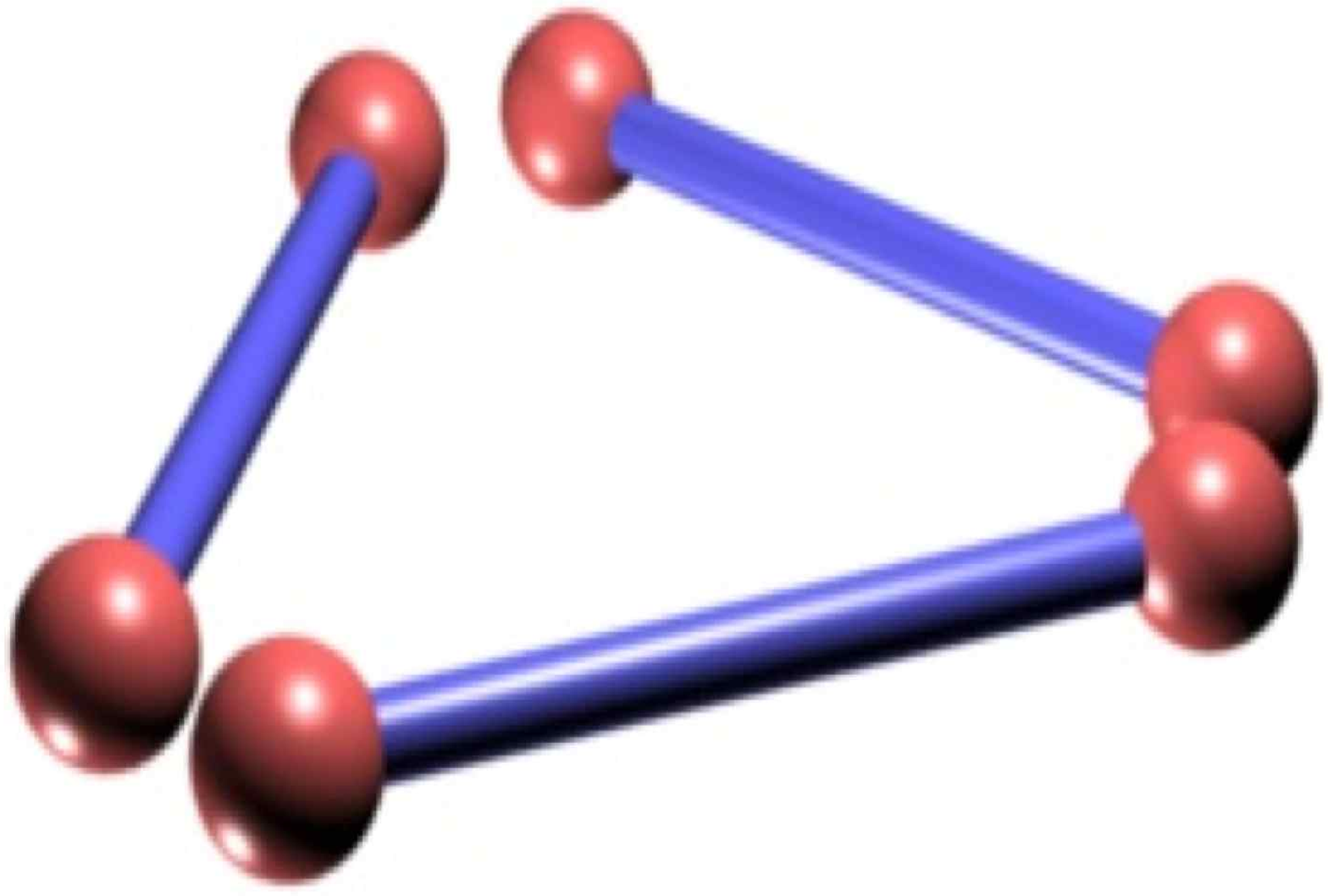}%

  \label{fig:hextria}
  \caption{Top row: A hexagonal and a triangular lattice. Each \bwtext{dark}{green} bond is constituted by two entangled pairs (here shown as \bwtext{light}{blue} bonds). Every second vertex is highlighted.
  The bottom row shows how the transformation works: Pairs of qubits are taken from the highlighted vertices. Onto the respective pairs of edges, the 
  protocol of Eq.\ (\ref{OptimRep}) is applied, succeeding with probability 
  $p$. Using this quantum operation,
  the lattice is transformed from a hexagonal lattice to a new triangular
  lattice (for cluster states
  this actually corresponds to performing a $\sigma_y$-measurement on 
  the highlighted sites). On this new triangular lattice, an edge is open if the protocol of
  Eq.\ (\ref{OptimRep}) has been successful. }
\end{figure}

We now turn to a strategy exploiting the more potential of the quantum
setting. We resolve every second site as shown in Fig.~\ref{fig:hextria}. Now -- as described in the caption of Fig.\ \ref{fig:hextria},
one performs the protocol of Eq.\ (\ref{OptimRep}), leading to a triangular
lattice. The edge probability in this new triangular lattice is given by
\begin{equation}
	2\lambda_2 = 2(1-\lambda_1)> 2 \sin(\pi/18)=p_{\Delta}^{\rm(c)}\approx
	0.3473
\end{equation}
with the above choice for $\lambda_1$. Hence we are now above the percolation
threshold for the triangular lattice, and one can proceed as usual to demonstrate
that an infinite cluster emerges. 
Therefore, bond percolation processes with quantum bonds are related to each other in a different way than those with classical bonds. More specifically, it is possible to find
a parameter regime for which the hexagonal lattice does not percolate,
but can be transformed into a percolating triangular lattice, if 
quantum operations are allowed for.

Equally interesting is the proposal to use 
the optimal singlet conversion strategy to transform a 
square lattice into two independent square lattices of 
doubled size, for which the bond probability is 
larger than in the original lattice.

\paragraph{Square lattice}

Another striking example of what can happen in quantum 
percolation is a nested protocol of distillation steps. This has been
proposed in Ref.\ \cite{PCAL07}, which develops the ideas of 
Ref.\ \cite{ACL07} in more detail.
Here, a set of non-maximally entangled pairs will be used which
cannot be transformed into maximally entangled ones deterministically by employing 
the process mentioned above, so $p_{1}<1$.

\begin{figure}[t]
  \psfrag{a}[bl][bl]{(a)}
  \psfrag{c}[bl][bl]{(b)}
  \psfrag{o}[bl][bl]{(c)}
  \psfrag{2}[bl][bl]{$|\varphi\rangle$}
  \psfrag{4}[tl][tl]{$|\varphi\rangle$}
  \psfrag{3}[tr][tr]{$|\varphi\rangle$}
  \psfrag{1}[br][br]{$|\varphi\rangle$}
  \psfrag{f}[bc][bc]{$|\varphi'\rangle$}
  \psfrag{g}[tc][tc]{$|\varphi'\rangle$}
  \psfrag{h}[bc][bc]{$|\varphi''\rangle$}
  \centering\includegraphics[width=.75\textwidth]{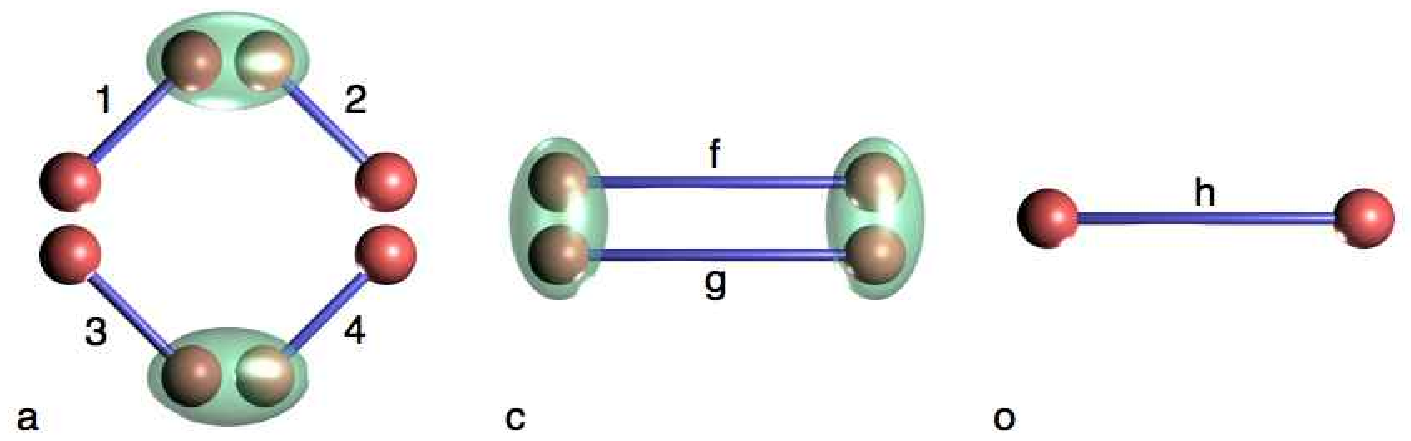}
  \caption{Transformation of a square lattice to another square lattice by means of a pair of (a) two-partite measurements and (b) an entanglement distillation on the resulting state.
    Starting with a square lattice where each bond consists of $|\varphi\rangle^{\otimes2}$ (similar to Fig.~\ref{fig:hextria}), this process generates a square lattices
    with $\sqrt{2}$ times the lattice constant.
    \label{fig:square} }
\end{figure}

Consider the situation of Fig.~\ref{fig:square}: Initially, arranged in a square,
one has four identical pairs in a state with state vector 
 $|\varphi\rangle$. Then one can perform local operations to the
 marked subsystems, as Bell-measurements in the 
 $\sigma_x\sigma_z$-basis (Fig.~\ref{fig:square}a). The original states $|\varphi\rangle$ have
Schmidt coefficients $\{\lambda_1,\lambda_2\}$, the new state 
with state vector 
$|\varphi'\rangle$ will have Schmidt coefficients, the smaller of which is 
given by
\begin{equation}
	\lambda_2'=(1-({1-(4\lambda_1\lambda_2)^2})^{1/2})/2.
\end{equation}
Given the four original input states, by applying this local filtering
twice, two such output states will be left. By using a further
local operation (Fig.~\ref{fig:square}b) --  an instance of a distillation protocol --
a final state vector $|\varphi''\rangle$ can be achieved (Fig.~\ref{fig:square}c) that is more 
entangled than $|\varphi'\rangle$.  The largest Schmidt coefficient of 
this final state is given by 
$\lambda_1''=\max\{1/2,(\lambda_1')^2\}$ \cite{PCAL07,NV01}.
This state is maximally entangled iff both Schmidt coefficients 
are equal, so iff $\lambda''_1=1/2$. 
The range in which that can be achieved is
\begin{equation}
 1/2 \le \lambda_1 \le \lambda_1^* = \frac{1}{2}\left(1+({1-({2(\sqrt{2}-1)})^{1/2}})^{1/2}\right) \approx 0.65 .
 \end{equation}
Hence, certain non-maximally entangled ``quantum edges'' (with CEP probability smaller than one)
can be deterministically transformed into maximally entangled bonds, so 
edges that are open with unit probability.
 
Note that each of the above steps involving local operations can be 
achieved deterministically, resorting to LOCC operations. In each case,
the measurement outcome on each vertex has to be available to the
other vertices to do the appropriate corrections and to render the 
scheme deterministic.
Such an idea can be used to transform a square lattice, in which vertices
share ``quantum edges'' with nonunit CEP bond probability in form of non-maximally entangled states
into a square lattice with unit edge probability.

\begin{theorem}[Singlet conversion in the square lattice]
  A square lattice where nearest neighbors share $|\varphi\rangle^{\otimes2}$ can be transformed into
  a square lattice, the edges of which being fully occupied by maximally entangled pairs, using
  collective measurements at the individual sites, given that $\lambda_1<\lambda_1^*$.
\end{theorem}
One could say that 
the protocol transforms a square lattice with
\begin{equation}
  1\ge p_{1}\ge 2\lambda_2^*\approx.70
\end{equation}
into a square lattice with $p_{1}=1$. Again, and more clearly so, 
this is a transformation between different bond percolation processes 
that crosses the critical threshold in a way that is 
not possible in percolation with classical edges.

\paragraph{Square lattice II}
\begin{figure}[t]
  \includegraphics[width=.48\textwidth]{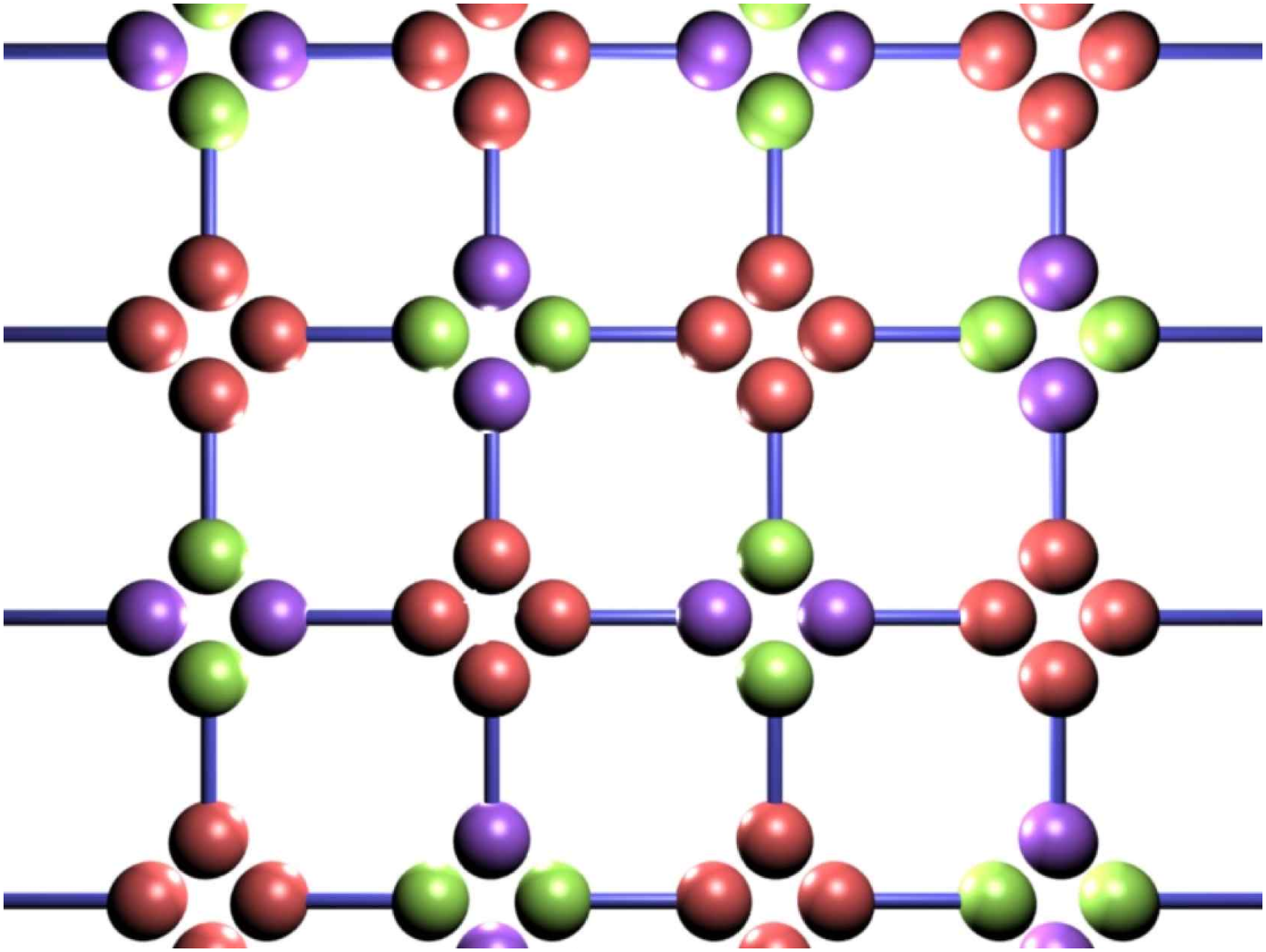}\hfill%
  \includegraphics[width=.48\textwidth]{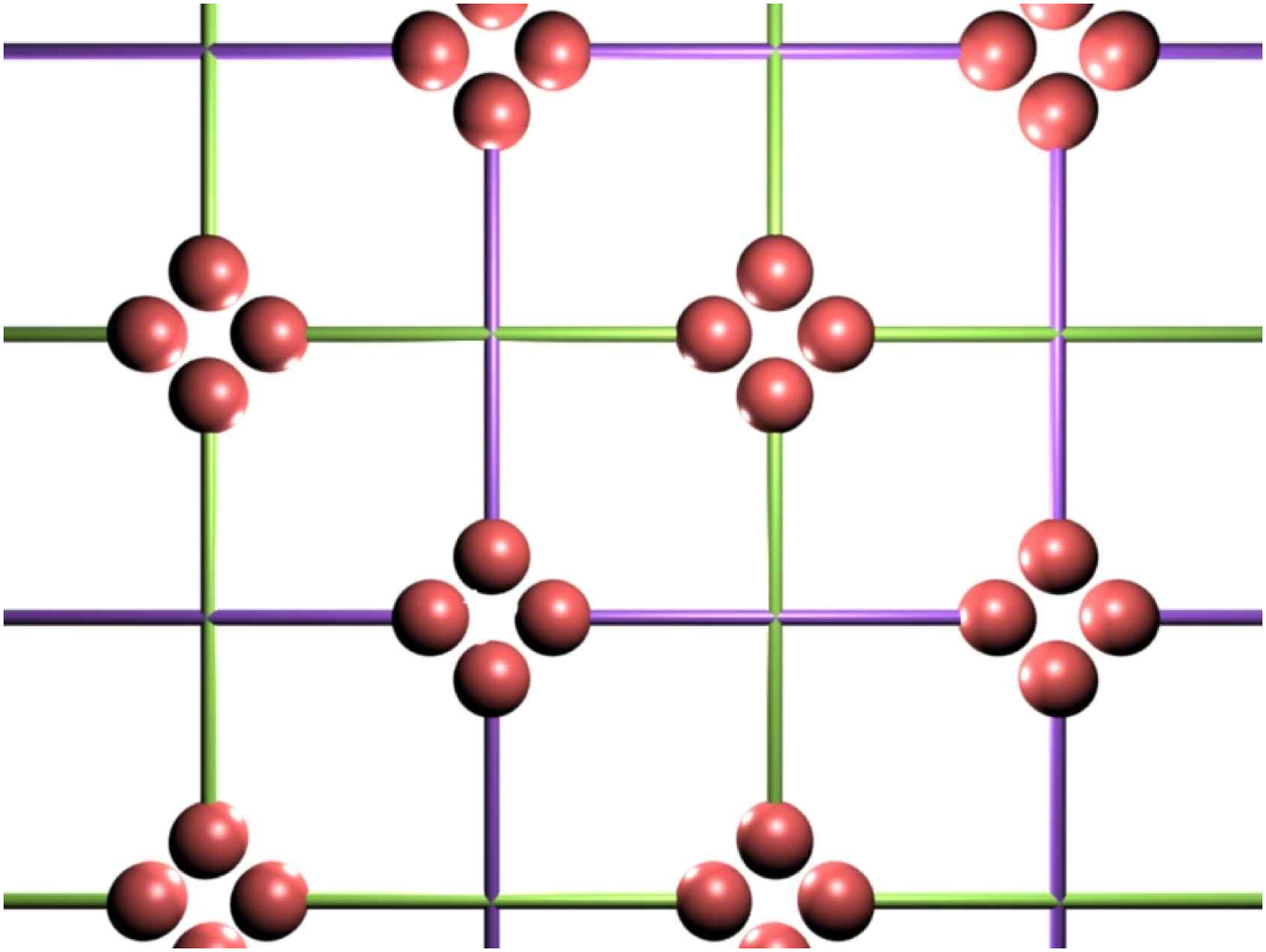}
  \caption{ Transformation of a square lattice into two disjoint ones by means of two qubit measurements. Left: At every other lattice site
    two operations are carried out, affecting the \bwtext{dark}{green} and \bwtext{light}{purple} qubits, respectively. Right: The resulting lattice consists of
    two disjoint square lattices. \label{fig:square2}}
\end{figure}

To consider yet another effect, we will have a second look at the square lattice\cite{PCAL07}.
As mentioned in Section~\ref{section:qperc}, a joint measurement
can be applied to a two-pair chain such that the SCP of the new
bond equals the original bonds' SCP on average.
Let us apply this operation at every other site of a square lattice
as indicated in Fig.~\ref{fig:square2}, thereby replacing it with two disjoint
square lattices with double the lattice constant.

Neighboring sites in the resulting graph sit on different lattices.
We pick two pairs of neighbors, $a=(A,A')$ and $b=(B,B')$ and ask for the probability
of finding at least one open path between these two sets of sites, so the event $\mathfrak{H}=a\leftrightarrow b$.
On the doubled lattices this accounts to the event of having at most one path out of
$A\leftrightarrow B$ and $A'\leftrightarrow B'$. In the limit of large separation of $a$ and $b$,
the probability amounts to
\begin{equation}
  P_p'(\mathfrak{H}) = \theta^2(p)(2-\theta^2(p)),
\end{equation}
where $\theta(p)$ is the percolation probability, so the probability of a given site to belong to an infinite spanning path.

In contrast, on the original lattice this quantity asymptotically fulfils
\begin{equation}
  P_p(\mathfrak{H}) \le \theta^2(p)(2-P_p(A\leftrightarrow A'))^2 .
\end{equation}
In Ref.~\cite{PCAL07} a Monte Carlo simulation to find the quantity $P_p(A\leftrightarrow A')$ for $p\searrow p_2^{\rm(c)}$ is
carried out, yielding $2-P_p(A\leftrightarrow A')\approx1.313$ in comparison to $(2-\vartheta^2(p_2^{\rm(c)}))^{1/2}=\sqrt{2}$.
So, a doubling of the lattice spacing using joint quantum operations can be advantageous when it comes to the connection of pairs of sites.
It is conjectured~\cite{PCAL07} that this property holds for all $p$.
Further examples along similar lines are also presented in Ref.\ \cite{PCAL07}.

\section{Summary and open problems}

\subsection{Open problems}

To elaborate the aforementioned invitation to this topic, 
we list some open
problems that arose from the work so far.
Concerning issues of 
``classical percolation for cluster state preparation'' 
this includes the following list of problems:
\begin{itemize}
  \item In dimensions $d\ge3$ crossing paths in different 
  directions do not imply
    a crossing of these paths. Further development could 
    aim at finding an algorithm
    similar to the one in Section~\ref{section:transition}. These would be able to identify
    structures that are topologically 
    equivalent to universal computing resource states
    in a percolated lattice at higher dimension, 
    thus allowing this tool to be used for
    resource state production for arbitrary gate probabilities with resources amounting to $O(L^2)$.
    
  \item While it appears from numerical results that a constant
    overhead per site cannot be achieved with 
    static renormalization, its optimal scaling
    is not evident. According to Fig.~\ref{fig:diamond_thresholds}, 
    this is at most
    logarithmic, rather than polynomial as the 
    bound in Theorem~\ref{thm:scaling} suggests.
    
  \item Further, an investigation towards the 
  required initial resource size
    would help implement these protocols in the future. 
    After all, small highly entangled
    quantum states have to be produced in the very beginning, 
    the smaller the better.
    There will be a tradeoff between the possible gate probabilities and
    the number of initial qubits needed, the exact behaviour of which might be interesting.
    It includes optimization over possible local linear optical operations which do not
    require global re-routing.
    
  \item All ideas presented in this chapter presume 
  ideal states, optical elements, or quantum gates.
    This, however, will of course never be achieved in any implementation. The general
    problem of quantum error correction needs to be revisited in the framework of
    percolation, as there might arise 
    new ways for error correction of special classes
    of errors (for first ideas see Ref.\ \cite{KRE06}).

\item
  As has been 
  shown in Ref.\ \cite{GE07}, there are a number of resource
  states very different from the cluster states which form universal 
  resources for quantum computing. Dictated by the underlying
  physical architecture, there might be states that are more
  suitable than cluster states, 
  exploiting the specific strengths of the respective architecture.
  However, if probabilistic processes are involved (like PEPS
  projections with linear optics), it is not obvious how to generalize
  the percolation scheme to other states.

%
  Information flow through these states is in general not ruled as
  easily as with single qubit Pauli measurements. Suitable
  applications of SWAP gates might help in using percolated versions
  of these resources for computing, though not giving a solution for
  distillation of hole-free states as such (as may be needed as resource
  for error correction).
  
\end{itemize}
Complementing these questions on the tightness of the
given bounds, there are a number of open questions
relating to the idea of  
entanglement percolation:
\begin{itemize}
  \item The aim of Section~\ref{section:networks} was to look at the problem of entangled state
    distribution with the tool of percolation. One -- relatively
    vague -- open problem is to find applications of this idea
    in a context different from entanglement distribution
    as such.

    \item One step in this direction could be to combine the idea
    of entanglement percolation with ones of quantum computation
    in the first part of this book chapter. Clearly, entanglement percolation could also help to further reduce the required resources
    in a renormalized lattice, albeit requiring more difficult 
    collective measurements on the way, the probabilistic nature of which
    might lead to some trade-off.
    
    \item Within restricted classes of percolation settings (such as
    restrictions to the geometry or the possible local operations), it
    seems important to identify the optimal entanglement
    percolation strategy.
    
    \item It is still far from clear how to fully incorporate mixed states
    in this setting, which seems important when considering lossy
	quantum channels.	
\end{itemize}

\subsection{Summary}

Even though the first ideas to incorporate quantum information 
with percolation theory were only proposed very recently
\cite{KRE06,ACL07},
it seems clear that open problems in 
quantum state preparation and quantum communication 
can indeed benefit from results known in percolation theory.
Since randomness is intrinsic in quantum mechanics,
one often has to overcome the probabilistic nature of 
quantum operations. As we have discussed, a 
context in which this most naturally is in quantum 
computation using lattice system. Here
static lattices with a non-unit probability of the
existence for bonds and sites occur naturally 
in cluster state preparation, where the random nature
is due to probabilistic quantum gates, or due to Mott
defects in the preparation. Ideas of percolation -- 
specifically questions of when crossing clusters exist --
led to methods of renormalization that effectively remove
this probabilistic aspect from quantum state preparation.
Along similar lines, it seems realistic to expect related
instances where percolation ideas help to overcome the 
intrinsic randomness of quantum mechanical state
manipulation. 

A key challenge will be, needless to say, to take mixed
quantum states fully into account, 
and to see in quantitative terms to what extent
renormalization ideas can be combined with methods
of fault tolerance, error correcting codes, or 
protection strategies in state preparation for the use 
of quantum computation \cite{DHM06,VBR06,VBR07,P97}. 
A step into this direction has been taken in Ref.\ \cite{KRE06}.

We have also seen that when taking the quantum nature of the
states seriously, one can often outperform strategies based
on simple measurements and invoking notions of edge 
percolation. Such an approach seems particularly suitable
for realizations in quantum networks for quantum communication
and key distribution, but could equally well also be applied to 
the above context of quantum computation. New phenomena
emerge when appropriate collective operations are allowed for,
giving rise to an interesting interplay between questions
of entanglement theory on the one hand and percolation theory 
on the other hand. 

On a related but different note, ideas of percolation seem 
also provide powerful tools in a slightly different context, namely
to study correlation and entanglement properties of quantum
many-body systems from the perspective of quantum
information. Notably, the scaling of the entanglement entropy
in the Ising model can be assessed by invoking concepts of 
classical percolation \cite{GOS07}. The question of relating
percolation ideas to problems in quantum information science
is at its infancy, but one should 
expect more applications to come.

\section*{Acknowledgments}
We warmly thank T.\ Rudolph for numerous fruitful and
fun discussions on the subject of the connection between
percolation theory and quantum information theory, who
is also coauthor of the joint publication Ref.\ \cite{KRE06}. We also
thank  T.\ Acin, D.E.\ Browne, W.\ D{\"u}r,   
A. Miyake, and T.J.\ Osborne for 
correspondence on percolation ideas in the
quantum information context. This work has been supported by 
Microsoft Research through the European PhD Programme, the 
EPSRC, the EU (QAP), and the EURYI award scheme.


\begin{thebibliography}{10}

\bibitem{ACL07}
A.~Ac{\'i}n, J.~I. Cirac, and M.~Lewenstein.
\newblock Entanglement percolation in quantum networks.
\newblock {\em Nature Physics}, 3:256, 2007.

\bibitem{Aizenman97}
M.~Aizenman.
\newblock On the number of incipient spanning clusters.
\newblock {\em Nuclear Physics B}, 485:551, 1997.

\bibitem{Bazant07}
M.~Z. Bazant.
\newblock Largest cluster in subcritical percolation.
\newblock {\em Phys. Rev. E}, 62:1660--1669, 2000.

\bibitem{BBPS96}
C.~Bennett, H.~Bernstein, S.~Popescu, and B.~Schumacher.
\newblock Concentrating partial entanglement by local operations.
\newblock {\em Phys. Rev. A}, 53:2046, 1996.

\bibitem{BondyMurty}
J.~A. Bondy and U.~S.~R. Murty.
\newblock {\em Graph theory with applications}.
\newblock Macmillan, 1976.

\bibitem{BVK99}
S.~Bose, V.~Vedral, and P.~L. Knight.
\newblock Purification via entanglement swapping and conserved entanglement.
\newblock {\em Physical Review A}, 60:194, 1999.

\bibitem{BR07}
S.~Bravyi and R.~Raussendorf.
\newblock Measurement-based quantum computation with the toric code states.
\newblock {\em Physical Review A}, 76:022304, 2007.

\bibitem{BR01}
H.~J. Briegel and R.~Raussendorf.
\newblock Persistent entanglement in arrays of interacting particles.
\newblock {\em Physical Review Letters}, 86:910, 2001.

\bibitem{BEF+07}
D.~E. Browne, M.~B. Elliott, S.~T. Flammia, S.~T. Merkel, A.~Miyake, and
  A.~Short.
\newblock Phase transition of computational power in the resource states for
  one-way quantum computation.
\newblock {\em arXiv:0709.1729}, 2007.

\bibitem{BR04}
D.~E. Browne and T.~Rudolph.
\newblock Resource-efficient linear optical quantum computation.
\newblock {\em Phys. Rev. Lett.}, 95:010501, 2005.

\bibitem{CL01}
J.~Calsamiglia and N.~L{\"u}tkenhaus.
\newblock Maximum efficiency of a linear-optical Bell-state analyzer.
\newblock {\em Applied Physics B}, 72:67, 2001.

\bibitem{DHM06}
C.~Dawson, H.~Haselgrove, and M.~Nielsen.
\newblock Noise thresholds for optical quantum computers.
\newblock {\em Phys. Rev. Lett.}, 96:020501, 2006.

\bibitem{Eisert04}
J.~Eisert.
\newblock Optimizing linear optics quantum gates.
\newblock {\em Physical Review Letters}, 95:040502, 2005.

\bibitem{EW04}
J.~Eisert and M.~M. Wolf.
\newblock {\em Quantum computing, in Handbook of innovative computing},
  chapter~8, page 253.
\newblock Springer, New York, 2004.

\bibitem{GRTZ02}
N.~Gisin, G.~Ribordy, W.~Tittel, and H.~Zbinden.
\newblock Quantum cryptography.
\newblock {\em Rev. Mod. Phys.}, 74:145, 2002.

\bibitem{Grimmett}
G.~Grimmett.
\newblock {\em Percolation}.
\newblock Springer, 2nd edition, 1999.

\bibitem{GOS07}
G.~Grimmett, T.~J. Osborne, and P.~Scudo.
\newblock Entanglement in the quantum ising model.
\newblock 2007.

\bibitem{GE07}
D.~Gross and J.~Eisert.
\newblock Novel schemes for measurement-based quantum computing.
\newblock {\em Physical Review Letters}, 98:220503, 2007.

\bibitem{GESP07}
D.~Gross, J.~Eisert, N.~Schuch, and D.~Perez-Garcia.
\newblock Measurement-based quantum computation beyond the one-way model.
\newblock {\em Physical Review A}, 76:052315, 2007.

\bibitem{GKE06}
D.~Gross, K.~Kieling, and J.~Eisert.
\newblock Potential and limits to cluster state quantum computing using
  probabilistic gates.
\newblock {\em Physical Review A}, 74:042343, 2006.

\bibitem{Hammersley}
J.~M. Hammersley and W.~Morton.
\newblock Poor man's Monte Carlo.
\newblock {\em Journal of the Royal Statistical Society (B)}, 16:23, 1954.

\bibitem{HDE+06}
M.~Hein, W.~D{\"u}r, J.~Eisert, R.~Raussendorf, M.~V. den Nest, and H.-J.
  Briegel.
\newblock Entanglement in graph states and its applications.
\newblock {\em quant-ph/0602096}, 2006.

\bibitem{HEB04}
M.~Hein, J.~Eisert, and H.~J. Briegel.
\newblock Multiparty entanglement in graph states.
\newblock {\em Physical Review A}, 69:062311, 2004.

\bibitem{HK76}
J.~Hoshen and R.~Kopelman.
\newblock Percolation and cluster distribution. {I}. cluster multiple labeling
  technique and critical concentration algorithm.
\newblock {\em Physical Review B}, 14:3438, 1976.

\bibitem{KGE06}
K.~Kieling, D.~Gross, and J.~Eisert.
\newblock Minimal resources for linear optical one-way computing.
\newblock {\em Journal of the Optical Society of America B}, 24(2):184, 2006.

\bibitem{KGE07}
K.~Kieling, D.~Gross, and J.~Eisert.
\newblock Cluster state preparation using gates operating at arbitrary success
  probabilities.
\newblock {\em New Journal of Physics}, 9:200, 2007.

\bibitem{KRE06}
K.~Kieling, T.~Rudolph, and J.~Eisert.
\newblock Percolation, renormalization, and quantum computing with
  non-deterministic gates.
\newblock {\em Phys. Rev. Lett.}, 99:130501, 2007.

\bibitem{KLM01}
E.~Knill, R.~Laflamme, and G.~Milburn.
\newblock A scheme for efficient linear optics quantum computation.
\newblock {\em Nature}, 409:46, 2001.

\bibitem{LBB+06}
Y.~L. Lim, S.~Barrett, A.~Beige, P.~Kok, and L.~Kwek.
\newblock Repeat-until-success quantum computing using stationary and flying
  qubits.
\newblock {\em Phys. Rev. A}, 73:012304, 2006.

\bibitem{MGW+03}
O.~Mandel, M.~Greiner, A.~Widera, T.~Rom, T.~W. H{\"a}nsch, and I.~Bloch.
\newblock Controlled collisions for multi-particle entanglement of optically
  trapped atoms.
\newblock {\em Nature}, 425:937, 2003.

\bibitem{nielsen}
M.~A. Nielsen and I.~Chuang.
\newblock {\em Quantum computation and quantum information}.
\newblock Cambridge University Press, 2000.

\bibitem{NV01}
M.~A. Nielsen and G.~Vidal.
\newblock Majorization and the interconversion of bipartite states.
\newblock {\em Quantum Information and Computation}, 1:76, 2001.

\bibitem{PCAL07}
S.~Perseguers, J.~I. Cirac, A.~Acin, M.~Lewenstein, and J.~Wehr.
\newblock Entanglement distribution in pure state quantum networks.
\newblock 2007.

\bibitem{PJF01}
T.~B. Pittman, B.~C. Jacobs, and J.~D. Franson.
\newblock Probabilistic quantum logic operations using polarizing beam
  splitters.
\newblock {\em Physical Review A}, 64:062311, 2001.

\bibitem{P97}
J.~Preskill.
\newblock Fault-tolerant quantum computation.
\newblock {\em quant-ph/9712048}, 1997.

\bibitem{RB01}
R.~Raussendorf and H.-J. Briegel.
\newblock A one-way quantum computer.
\newblock {\em Physical Review Letters}, 86:5188, 2001.

\bibitem{RBB03}
R.~Raussendorf, D.~E. Browne, and H.~Briegel.
\newblock Measurement-based quantum computation with cluster states.
\newblock {\em Physical Review A}, 68:022312, 2003.

\bibitem{VBR07}
R.~Raussendorf and J.~Harrington.
\newblock Fault-tolerant quantum computation with high threshold in two
  dimensions.
\newblock {\em Phys. Rev. Lett.}, 98:190504, 2007.

\bibitem{SL04}
S.~Scheel and N.~L{\"u}tkenhaus.
\newblock Upper bounds on success probabilities in linear optics.
\newblock {\em New Journal of Physics}, 6:51, 2004.

\bibitem{SW02}
D.~Schlingemann and R.~F. Werner.
\newblock Quantum error-correcting codes associated with graphs.
\newblock {\em Physical Review A}, 65:012308, 2002.

\bibitem{sedgewick}
R.~Sedgewick.
\newblock {\em Algorithms}.
\newblock Addison-Wesley, 1983.

\bibitem{NDM04}
M.~van~den Nest, J.~Dehaene, and B.~D. Moor.
\newblock Graphical description of the action of local clifford transformations
  on graph states.
\newblock {\em Physical Review A}, 69:022316, 2004.

\bibitem{NMDB06}
M.~van~den Nest, A.~Miyake, W.~D{\"u}r, and H.~J. Briegel.
\newblock Universal resources for measurement--based quantum computation.
\newblock {\em Phys. Rev. Lett.}, 97:150504, 2006.

\bibitem{VBR06}
M.~Varnava, D.~Browne, and T.~Rudolph.
\newblock Loss tolerance in one-way quantum computation via counterfactual
  error correction.
\newblock {\em Phys. Rev. Lett.}, 97:120501, 2006.

\bibitem{Vidal99}
G.~Vidal.
\newblock Entanglement of pure states for a single copy.
\newblock {\em Physical Review Letters}, 83:1046, 1999.

\bibitem{WRR+05}
P.~Walther, K.~Resch, T.~Rudolph, E.~Schenck, H.~Weinfurter, V.~Vedral,
  M.~Aspelmeyer, and A.~Zeilinger.
\newblock Experimental one-way quantum computing.
\newblock {\em Nature}, 434:169, 2006.

\end{thebibliography}
\end{document}